\documentclass[%
reprint,
 amsmath,amssymb,
 aps, physrev,
]{revtex4-2}

\usepackage{graphicx}
\usepackage{dcolumn}
\usepackage{bm}
\usepackage{xcolor}
\usepackage{hyperref}

\newcommand{\Mod}[1]{\ (\mathrm{mod}\ #1)}
\newcommand{\unit}{1\!\!1}

\usepackage{color}

\def\black#1{{\color{black} {#1}}}
\newcommand{\ket}[1]{|#1\rangle}

\begin{document}

\preprint{}

\title{\textbf{Lattice Quantization of Free Fermions without Doublers} 
}%

\author{Mario A. Serna~Jr.}
\affiliation{%
Physics Department, New Mexico Institute of Mining and Technology, Socorro, NM, USA}
\email{Contact author: mario.serna@nmt.edu}
\altaffiliation[Also a visitor at]{
CQuIC \& COSMIAC Research Centers, University of New Mexico, Albuquerque, NM, USA
}%

\author{Paul M.~Alsing}
\affiliation{
Department of Physics and Astronomy,
Florida Atlantic University, 
Boca Raton, FL, 
USA
}%


\date{\today}

\begin{abstract}
We present  a method to quantize free fermions which eliminates the doublers when implemented on the lattice in any number of dimensions and in the $m=0$ limit. 
The elimination of doublers is achieved by combining a second-order description of fermions, with the tools associated with non-Hermitian Hamiltonians.  We identify a new Pseudo-Hermitian symmetry of the second-order fermion equation, and we identify the associated $U(1)$ \black{symmetry} which will become charge when shifted to a local gauge theory.  We validated the methods numerically.
\end{abstract}

\maketitle


\section{\label{sec:Intro}Introduction}

\begin{figure*}
    \centering
    (a) \includegraphics[width=0.45\linewidth]{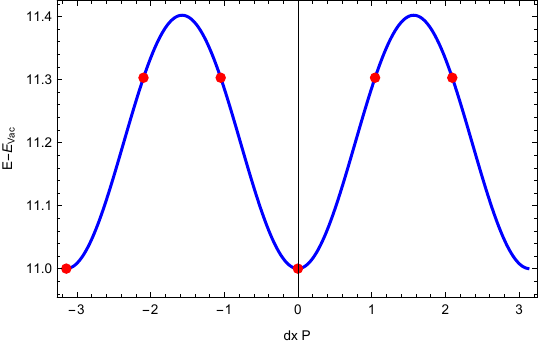}
    (b) \includegraphics[width=0.45\linewidth]{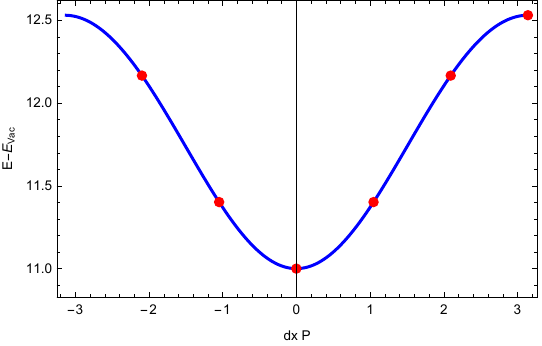}
    \caption{ 
  The `fermion doubling problem' is seen by comparing the dispersion relation for the first-order Dirac equation in (a) and the desired dispersion relation in (b) above. 
    The blue, continuous curves shows the theoretical dispersion relationships for single particle states.  
    The six, red dots overlaid show the numerical results for a $1+1D$ simulation with $6$ spatial points.
    The points in (a) with non-standard dispersion relations ($P>0$, $\frac{dE}{dP}<0$) and($P<0$, $\frac{dE}{dP}>0$) are the extra doubled states.  
    The red dots in (b) show the results of the technique described in this paper.
    The cases shown in (a) and (b) have $m=11$, $dx=1/3$. 
      \label{Fig:FirstOrderwSecondOrderDispersion} 
      }
\end{figure*}

Modeling gauge theories on future quantum computers is most natural in the lattice Hamiltonian formulation \cite{davoudi2025tasi}.
However, the quantization of fermions on a lattice suffers from a doubling of the spectrum known as the `fermion doubling problem' \cite{Kogut.PhysRevD.11.395,Tong2018}.
The core of the problem is that the first-derivative structure of the Dirac equation $(\gamma^\mu \partial_\mu - m )\psi =0$ yields the dispersion relation shown in the blue, continuous curve in Fig.~\ref{Fig:FirstOrderwSecondOrderDispersion} (a). 
The numerical, first-derivative $\partial_x \psi \equiv \frac{1}{2\,dx} (\psi(x+dx) - \psi(x-dx))$ skips a lattice site and therefore aliasing causes a doubling of the expected spectrum. 
The six, red dots overlaid in Fig.~\ref{Fig:FirstOrderwSecondOrderDispersion} (a) show the numerical results for a simulation of the $1+1$D Dirac equation with $6$ spatial points.
One can see the spectrum doubling: there are two solutions with the same $E$ for $P>0$, and two solutions for the same $E$ for $P<0$.
The points in Fig.~\ref{Fig:FirstOrderwSecondOrderDispersion}(a) with non-standard dispersion relations ($P>0$, $\frac{dE}{dP}<0$) and($P<0$, $\frac{dE}{dP}>0$) are the extra doubled states.  
Fig.~\ref{Fig:FirstOrderwSecondOrderDispersion} (b) shows the desired dispersion relationship.
As the number of points goes to infinity and the lattice spacing ($dx$) goes to zero, the dispersion spectrum approaches the continuum theory.

In this paper we present an alternative technique based on a second-order model of fermions on the lattice.
The red dots in Fig.~\ref{Fig:FirstOrderwSecondOrderDispersion} (b) show the results of a numerical simulation of our technique. 
These results match the desired dispersion relationship where the doublers have disappeared.
This approach circumvents the Nielsen Ninomiya no-go theorem \cite{NIELSEN198120} by being second-order.
The technique works for multi-dimensional spatial lattices,
\black{as well as}
in the limit where $m=0$.
Because there are no doublers, the same computational resources can better model the various energy and momentum values.

There are three novel features to this technique:  
First we identify the $U(1)$ symmetry present in a second-order real Majorana fermion model that corresponds to the $U(1)$ \black{symmetry} in the first-order formulation.
Second, we identify a new pseudo-Hermitian symmetry $\eta$ specific to this model.
Third, we demonstrate that when quantized on a lattice the fermion doublers disappear in this system.

Our discussion  begins in Sec.~\ref{sec:Background} with the background associated with second-order fermions and non-Hermitian quantum mechanics.
In Sec.~\ref{sec:SecondOrderFermionQuantization}, we begin with the first-order Dirac equation and show how to derive and canonically quantize the equivalent second-order fermion theory. 
Sec.~\ref{sec:NumericalResults} describes the numerical implementation of the quantization on a lattice.
We then compare the first-order and second-order cases, discuss the implications, and conclude in  Sec.~\ref{sec:Discussion}.
Last, in our Appendix, we provide added details on the various numerical tools required for the implementation of the technique.
     
\section{\label{sec:Background}Background}

The promise of quantum computers has renewed interest in the Hamiltonian formulation of lattice gauge theory \cite{davoudi2025tasi}.
However, the fermion doubling problem blocks one from having a viable lattice formulation for the electroweak theory that could then be simulated on a quantum computer.
The fermion doubling problem has been known for over 50 years \cite{Kogut.PhysRevD.11.395}.
In 1981, Nielson and Ninomiya  introduced a no-go theorem showing the doublers are an unavoidable, topological artifact of quantizing the Dirac Lagrangian \cite{NIELSEN198120}.
Some \black{first-order} approaches to work around it include the Kogut-Susskind staggered-fermion method \cite{Kogut.PhysRevD.11.395}, Wilson fermions \cite{PhysRevD.10.2445}, 
Domain-Wall fermions, and Overlap fermions \cite{kaplan2009chiral}. 
None fully resolves the issue \cite{davoudi2025tasi, kaplan2009chiral}.

The use of a second-order fermion equation evades the Nielsen and Ninomiya no-go theorem.
In the discussion of their no-go theorem, Nielsen and Ninomiya cite Banks and Casher's work on second-order fermions \cite{banks1980chiral} as a \black{possible} way to circumvent their assumptions. 
A numerical second-order derivative $\partial^2_x \psi = \frac{1}{2\,dx}(\psi(x+dx)+\psi(x-dx) - 2\psi(x))$ does not skip a lattice site and therefore does not have the same aliasing effects that cause the doubling.

Gell-mann and Feynman introduced a second-order fermion model in 1958 \cite{PhysRev.109.193}.
That same year, Kibble showed that canonical quantization of higher-dimensional Lagrangians associated with their second-order fermions  led to negative norm states \cite{kibble1958higher}.
In addition, the secon-order fermion models were shown to have non-Hermitian Hamiltonians \cite{espin2015second}.
Others have attempted to use second-order fermion models to avoid the fermion doubling problem, but the complexities associated with its quantization have prevented success \cite{petroni1985second,cufaro1988second,cortes1993second,palumbo1991second}.

Hope is found in the path-integral quantizations for second-order fermions. Perturbative approaches based on the Poincare projectors have been successful, and they reproduced many QED renormalization group  results \cite{Delgado_Acosta_2011,PhysRevD.85.076004}.
In 2013, Espin also developed a second-order fermion formulation of the standard model \cite{espin2013second}, and he identified a canonical transformation such that the quantization of the second-order system is equal to the first-order quantization \cite{espin2015second}. 
Sadly, this mapping just recreates the fermion-doubling problem.
In this paper, we show a distinct quantization.

In parallel in the early 2000s, Bender and Mostafazadeh discovered new techniques for working with non-Hermitian Hamiltonians which eliminate the negative-norm states \cite{bender2007making, Mostafazadeh_2002}. 
One identifies a pseudo-Hermitian symmetry $\eta$ of the Hamiltonian such that 
$H^\# \equiv \eta^{-1}\, H^\dag \, \eta = H$.
Then the inner product $\langle \Phi | \eta |\Phi'\rangle$ is conserved under time evolution, and the probability interpretation can be recovered.  The use of the pseudo-Hermitian conjugate $\#$ is to distinguish it from the more traditional Hermitian adjoint $\dag$.

More recently in 2024, Ferro, Olmos, Peinado, Vaquera (FOPV)  \cite{ferro2024quantization} showed one approach to canonically quantize a second-order free fermion theory with a global $U(1)$ symmetry, but their model corresponded to 8-spinor fermions with states beyond what one would expect for a basic electron. 

In the next section, we show an alternative quantization to FOPV corresponding to a more traditional $4$-spinor.
We identify a different symmetry $\eta$ and a different $U(1)$ symmetry which allows one to identify the states as those of a traditional QED electron.

\section{\label{sec:SecondOrderFermionQuantization} Second-Order Fermion Quantization in the Continuum}

We begin as did Feynmann and Gell-mann \cite{PhysRev.109.193} with the Dirac equation of motion
  \begin{equation}
      ( i\, \gamma^\mu D_\mu  - m )\psi = 0.
  \label{EqDiracEqMath}
   \end{equation}
where $D_\mu = \partial_\mu - i \textbf{A}_\mu$.
We absorb the charge $e$ and any associated generator into a redefinition of the vector potential $\textbf{A}_\mu$.
Following \cite{PhysRev.109.193}, we can define the Dirac \black{4-spinor} field  $\psi$ as being proportional to 
 \begin{equation}
    \psi \propto (i\, \gamma^\nu D_\nu  + m) \chi
\label{EqPsiFromChi2ndOrder}
 \end{equation}
where $\chi$ is a new Grassmann field.
Combining these gives a second-order  equation of motion for $\chi$
\begin{equation}
  \left(     D_\mu D^\mu + S^{\mu\nu} \, \textbf{F}_{\mu \nu} - m^2 \right) \chi = 0
  \label{EqSecondOrderEOMwA}
\end{equation} 
where $\textbf{F}_{\mu\nu} = \partial_\mu \textbf{A}_\nu - \partial_\nu \textbf{A}_\mu = i [D_\mu,D_\nu] $ and $S^{\mu\nu} = \frac{-i}{4} [ \gamma^\mu,\gamma^\nu]$, and we use a mostly negative metric.
Because the equation of motion for $\chi$ is second-order, there are twice as many solutions for it compared to the first-order $\psi$ for the Dirac field. 
For $\chi$ there is a solution associated with each set of initial values for $\chi$ and the derivative of $\chi$.
The dominant approach in the literature has been to cut the number of  $\chi$ fields in half by projecting $\chi$ onto one chiral representation $\chi=\frac{1}{2}(1 + \gamma^5)\chi$ \cite{PhysRev.109.193,espin2015second}.
We do not use this approach.

In contrast, we instead cut the number of fields in half by looking at $\chi$ as a real, Majorana, Grassmann field.
The Dirac field $\psi$ is complex and is defined by $8$ real Grassmann fields.
Therefore we should be able to represent the same dynamics with $4$ real Grassmann fields $\chi$ and the four derivatives of $\chi$.  

Because we are working with real Grassmann fields $\chi$,  we must work with Majorana representations of the gamma matrices, $\tilde \gamma^\mu$, and denote these special representations with a tilde following Pal's convention \cite{pal2011dirac}. 
In particular in the Majorana representation $\tilde \gamma^\mu$ are all imaginary and $\tilde \gamma^0 = -(\tilde \gamma^0)^T$ is antisymmetric so that terms in the Lagrangian like $\chi \tilde \gamma^0 \chi$ do not vanish.

The equations of motion in Eq.~\ref{EqSecondOrderEOMwA} follow from the Lagrangian density
\begin{eqnarray}
     {\mathcal{L}} & = & \frac{1}{2} (\tilde \gamma^\mu D_\mu \chi)_a (\tilde \gamma^0)^{ab}  (\tilde \gamma^\nu D_\nu \chi)_b
    - \frac{1}{2} m^2 \chi_a (\tilde \gamma^0)^{ab} \chi_b \\
& = & 
      \frac{-1}{2} \partial_\mu \chi\ \tilde \gamma^0 \partial^\mu \chi   - \frac{1}{2} m^2 \chi \tilde \gamma^0 \chi 
    +
  \frac{-i}{2} A_\mu [ \chi, \partial^\mu \chi ]  \nonumber
  \\
   & &
  + \frac{1}{4}F_{\mu\nu} \, \chi \,\left(  S^{\mu\nu} +  \tilde \gamma^0 S^{\mu\nu} \tilde \gamma^0 \right)\,\chi 
  \nonumber \\
 &  & -  A_\mu (\partial_\nu \chi) \left( S^{\mu\nu}  - \tilde \gamma^0 S^{\mu\nu}   
    \tilde{\gamma}^0 
     \right) \chi \nonumber \\  
&  &   -  \frac{1}{2} A_\mu \, A^\mu\,\chi \tilde{\gamma}^0  \chi .
\label{EqActionMajoranaRepSecondOrderWelectromagnetism}
\end{eqnarray}
On the first line for clarity, we have included the spinor indices $a$ and $b$.  
Because we are in a real representation, there is no dotted index on $\tilde \gamma^0$.
Although the $\chi$ field is real, there is a local $U(1)$ symmetry generated by $\tilde \gamma^0$.  
The vector field   $\textbf{A}_\mu = \tilde \gamma^0 A_\mu$ and the field strength tensor $\textbf{F}_{\mu\nu} = \tilde{\gamma}^0 F_{\mu\nu}$ 
have been identified with the generator of the local $U(1)$ symmetry $\tilde \gamma^0$.
The local symmetry sends
$\chi \rightarrow \exp(i\,\theta(x) \tilde \gamma^0) \chi$ and 
$\tilde \gamma^0 A_\mu \rightarrow \tilde \gamma^0 A_\mu - i \tilde \gamma^0 \partial_\mu \theta(x)$.
Last, to give the Lagrangian action units,  $\chi$ has units of $dx^{-(d-1)/2}$ where $d$ is the number of spatial dimension, and where $dx$ is the length scale for units in the theory. Once we discretize, $dx$ will be the lattice spacing.

Because our purpose is to introduce a new approach to quantizing second-order fermions which will avoid the doublers, we simplify to a global $U(1)$ symmetry and quantize the free fermion theory. 
Second-quantizing the discrete interacting theory will be the topic of a future publication.

The free, second-order Lagrangian density for the remainder of this work is 
\begin{equation}
    {\mathcal{L}} =  
    \frac{-1}{2} \partial_\mu \chi\ \tilde \gamma^0 \partial^\mu \chi   - \frac{1}{2} m^2 \chi \tilde \gamma^0 \chi .
\end{equation}
We select a time slice and canonically quantize.
The canonical conjugate momentum is
\begin{eqnarray}
    \pi^a_{\chi} & = & \frac{\delta^L {\mathcal{L}}}{\delta (\partial_0 \chi_a)}  = (\tilde \gamma^0)^{ab} \partial^0 \chi_b
\end{eqnarray}
and the associated Hamiltonian density is
\begin{equation}
    {\mathcal{H}} = 
     \frac{-1}{2} (\pi_\chi)^a  (\tilde \gamma^0)_{ac}  (\pi_\chi)^c   
      - \frac{1}{2}  \chi_a (\tilde \gamma^0)^{ab} ( \nabla^2 -m^2) \chi_b. 
      \label{EqHamiltonianContinuum}
\end{equation}
Based on the Poisson brackets \cite{henneaux1992quantization}, 
the quantum operators are defined by their equal-time, canonical quantization conditions :
\begin{eqnarray}
 \{ \chi_a (\vec x), \chi_b (\vec y) \} & \equiv & 0  \ \ \ \
 \{ \pi_\chi^a(\vec x), \pi_\chi^b(\vec y) \}  \equiv  0  \nonumber \\ 
 \{ \pi_\chi^a(\vec x), \chi_b (\vec y) \} & \equiv & -i \delta^a_b \delta^3(\vec x-\vec y) . \label{Eq2ndQuantizationMajoranaConditions}
\end{eqnarray}
If we had instead used complex fields as in previous quantizations of second order fermions \cite{kibble1958higher, ferro2024quantization}, one of the quantization relations would have been $\{ \chi^\dag , \chi \} = 0$ which requires the introduction of an auxiliary fields with a negative norm.
We instead quantize a purely real Grassmann field.

The second-order Hamiltonian is the spatial integral of Eq.~\ref{EqHamiltonianContinuum}, $H=\int d^3x {\mathcal{H}}$. This operator $H$ is then used to evolve the wave function $|\Psi\rangle$ through the first-order Schrodinger equation:
\begin{equation}
    i\,\partial_t |\Psi\rangle = H\, |\Psi \rangle. 
\end{equation}
To implement this quantization approach,  we look to inspiration from fermion wave functionals.
Unlike the first-order case, in the second-order formulism, the canonical momentum $\pi_\chi$ and the field $\chi$ have different units.
We shift basis to improve our intuition.
Following Floreanini and Jackiw \cite{floreanini1988functional},
we use as a basis Grassmann fields  $c_a(x)$ and $c^{\dag\,a}(x) \equiv \delta/\delta c_a(x)$ such that $c$ and $c^\dag$ have the same units.
The operators $c$ and $c^\dag$ satisfy $\{ c_a(x), c^{\dag b}(x')  \} = \delta^b_a \delta(x-x')$.
For $3+1$D, $c_a(x)$ has units of $dx^{-3/2}$.  The quantization conditions Eq.~\ref{Eq2ndQuantizationMajoranaConditions} are satisfied if we define $\chi_a(\vec x) \equiv \sqrt{dx}\,c_a(\vec x)$ and $\pi_\chi^a(x) = -i\,c^{\dag\,a}(x)/\sqrt{dx}$ where  $dx$ is a length scale that we can take as the lattice spacing.
This definition of Hermitian conjugate makes clear that the Hamiltonian density in Eq.~\ref{EqHamiltonianContinuum} is not Hermitian
\black{since 
$\chi_a^\dag\propto \pi_\chi^a$, and 
$(\pi_\chi^a)^\dag\propto \chi^a$,  therefore the Hermitian conjugation of $H$ effectively swaps the terms in the Hamiltonian Eq.~\ref{EqHamiltonianContinuum} but not the coefficients, so  ${\mathcal{H}}^\dag \ne {\mathcal{H}}$.
} 

Next we shift to momentum space with the convention
$   \tilde \chi_a (\vec p)  =  \int d^3 x  \, e^{-i\,\vec p \cdot \vec x} \chi_a(\vec x)$ 
and $    (\tilde{\pi}_\chi)^a (\vec p)  =  \int d^3 x  \, e^{-i\,\vec p \cdot \vec x} \pi_\chi^a(\vec x)$.
Because $\chi_a(\vec{x})$ is a real Grassmann field,  the complex conjugate of the Fourier transformed field satisfies: $\tilde{\chi}_a(\vec{p})=\tilde{\chi}^*_a(-\vec{p})$ and $\tilde{\pi}^a_\chi(\vec{p})=\tilde{\pi}_\chi^{*a}(-\vec{p})$.
In momentum space $\tilde{\chi}$ has units of $dx^2$ and $\tilde{\pi}_\chi$ has units of $dx$.
The Hamiltonian is now
\begin{eqnarray}
  H &=&   \frac{-1}{2} \int \frac{d^3\vec{p}}{(2\pi)^3}
     \left( \tilde \pi^*_\chi(\vec{p})  \tilde \gamma^0  \tilde \pi_\chi(\vec{p})   
      -  \omega^2_p\, \chi^*(\vec{p}) \tilde \gamma^0  \tilde{\chi}(\vec{p}) 
      \right) \hspace{0.5cm}
\label{EqHamiltonianContinuumMomentumSpace}
\end{eqnarray}
where $\omega_p = +\sqrt{\vec{p}^{\,2} +m^2}$.
The equal-time, quantization conditions in momentum space are
\begin{eqnarray}
 \{ \tilde{\pi}_\chi^{*c}(\vec{p}), \tilde{\chi}_b(\vec{p}\,') \}  & = &  - i  \delta^c_b (2\pi)^3 \delta^3(\vec p-\vec{p}\,') 
 \\ 
    \{ \tilde{\pi}_\chi^{c}(\vec p), \tilde{\chi}^*_b(\vec{p}\,') \}  & = & - i  \delta^c_b (2\pi)^3 \delta^3(\vec p-\vec{p}\,') 
\label{EqComlexModeCommutationRelations}.
\end{eqnarray}
Because $\chi(\vec{x})$ is real, the equal-time quantization conditions can be equated to $\{ \tilde{\pi}_\chi(\vec{p}), \tilde{\chi}(\vec{p}\,' ) \} = -i (2\pi)^3 \delta^3(\vec{p}+\vec{p}\,')$.
All other combinations anticommute to $0$.
From the Fourier transform and the anticommutation relations, we have: $\tilde \chi_a (\vec p) = \sqrt{dx} \, \tilde{c}_a (\vec p)$ and 
$\tilde{\pi}_\chi^{* b}(\vec p) = -i\,\frac{\delta}{  \delta \tilde{\chi}_b(\vec p)} =\frac{-i}{\sqrt{dx}} \tilde{c}^{\dag\,*\,b} (\vec p)$.
The momentum space $\tilde{c}$ and $\tilde{c}^\dag$ operators satisfy
$\{ \tilde{c}_a(\vec{p}), \tilde{c}^{\dag *\,b}(\vec{p}\,') \} = \delta_a^b (2\pi)^3 \delta^3(\vec p - \vec p ')$
.
This implements the canonical anitcommutation relations in  Fourier space.

The Hamiltonian is pseudo-Hermitian and satisfies
$H^\# = \eta^{-1} H^\dag \eta = H$ if
\begin{eqnarray}
    (\tilde \pi_\chi^a (\vec{p}))^\# & = &  \eta^{-1} (\tilde \pi_\chi^a(\vec{p}))^\dag \, \eta = i \, \omega_p \, \tilde \chi^*_a (\vec{p}) \\
  (\tilde \chi_a (\vec{p}))^\# & = &  \eta^{-1} (\tilde{\chi}_a(\vec{p}))^\dag \, \eta =  \frac{i}{\omega_p\,} (\tilde \pi_\chi^{*a} (\vec{p})) . 
\end{eqnarray}
Re-expressed in terms of the $c$ and $c^\dag$ operators, the requirement is:
\begin{eqnarray}
    (\tilde \pi_\chi^a (\vec{p}))^\# & = &  \eta^{-1} \frac{(\tilde{c}^{\dag\,a}(\vec{p}))^\dag }{\sqrt{dx}}\, \eta =   \omega_p \, \sqrt{dx}\, \tilde{c}^*_a (\vec{p}) 
   \label{EqPiSymmetry} \\
  (\tilde \chi_a (\vec{p}))^\# & = &  \eta^{-1} \sqrt{dx}\, (\tilde{c}_a(\vec{p}))^\dag \, \eta =  \frac{1}{\omega_p\,} \frac{(\tilde{c}^{\dag \,* \, a} (\vec{p}))}{\sqrt{dx}} . 
  \label{EqChiSymmetry}
\end{eqnarray}
This system can be solved (see Appendix \ref{SecEtaSolution}) for the operator $\eta$ yielding 
\begin{equation}
   \eta \equiv \exp\left( - \int \frac{d^3\vec{q}}{(2\pi)^3} \,\log( \omega_{q} dx) \,\tilde{c}^*_a(\vec q) \, \tilde{c}^{\dag a} (\vec{q}) \right).
   \label{EqEtaDef}
\end{equation}
The $\eta$ solution has a simple interpretation.
For every mode $\vec{p}$ there are four degrees of freedom associated \black{with $\pm q$ for each two state (qubit)} spinor index of  $\tilde{\chi}_a(\vec{p})$.
The operator in the exponent ( $\,\tilde{c}^*_a(\vec q) \,\tilde{c}^{\dag a} (\vec{q})$ ) is a counting operator for the \black{$0$-state} qubits of the four states with momentum \black{ $\pm\vec q$}.
Therefore the operator $\eta$  effectively normalizes the state by a factor $(dx\,\omega_q)^{-1}$ for each of these qubit states for that mode that is not occupied.

\black{We now} build the ladder operators of definite charge. 
States of definite charge are  eigenvectors of the $U(1)$ generator $\tilde \gamma^0$.
The $\tilde \gamma^0$ matrix performs many roles in our theory.  It raises or lowers spin indices and therefore can have both indices raised or lowered.  
However, when acting as part of a Lorentz generator or as the generator of the $U(1)$ symmetry, it has a raised and a lowered index.

To conform with traditional notation, let the eigenvectors associated with $+1$ and $-1$ eigenvalues of $\tilde \gamma^0$ be denoted as $u$ and $v$ respectively:
\begin{eqnarray}
    (\tilde \gamma^0)^{a}_{\ b} u^b_s & = & u^a_s ,\\
    (\tilde \gamma^0)^{a}_{\ b} v^b_s & = & -v^a_s,
\end{eqnarray}
where $s$ indexes the two spin states $0$ to $1$.  
We normalize these eigenvectors as $(u^\dag)_s u^r = \delta_s^{r}$  and $(v^\dag)_s v^r = \delta_s^{r}$  and $(v^\dag)_s\, u^r=0$.
Unlike the first-order case, $u$ and $v$ do not depend on $\vec{p}$.
The components of $u$ and $v$ also form the unitary matrix  that converts the Majorana-basis gamma matrices $\tilde \gamma^\mu$ to the standard basis where $\gamma^0_S$ is diagonal.

The lowering operators for positive charges are
\begin{eqnarray}
   a_s(\vec p) & = & u^c_s\,\sqrt{\frac{\omega_p}{2}}   \left( \tilde{\chi}_c(\vec p) + i \frac{1}{\omega_p} (\gamma^0)_{cb} \tilde{\pi}^b_\chi(\vec p) \right),
\end{eqnarray}
and the raising operators for positive charge are
\begin{eqnarray}
    (a^\#)^s(\vec p) & = & \eta^{-1}\, (a_s(\vec p))^\dag \eta  \\
    & = & (u^\dag)^s_c \sqrt{\frac{\omega_p}{2}}\left( \tilde{\chi}^*_{b}(\vec p)\, (\tilde \gamma^0)^{bc} + \frac{i}{\omega_p} \tilde{\pi}_\chi^{*\,c}(\vec p)
    \right). \nonumber 
\end{eqnarray}
The lowering ladder operators for negative charge are
\begin{eqnarray}
   b_s(\vec p) 
   & = & v^c_s\,\sqrt{\frac{\omega_p}{2}}   \left( \tilde{\chi}_c(\vec p) + i \frac{1}{\omega_p} (\gamma^0)_{cb} \tilde{\pi}^b_\chi(\vec p) \right),
\end{eqnarray}
and a raising operator
\begin{eqnarray}
    (b^\#)^s(\vec p) & = & \eta^{-1}\, (b_s(\vec p))^\dag \eta  \\ 
    & = & (v^\dag)^s_c \sqrt{\frac{\omega_p}{2}}\left( \tilde{\chi}^*_{b}(\vec p)\, (\tilde \gamma^0)^{bc} + \frac{i}{\omega_p} \tilde{\pi}_\chi^{*\,c}(\vec p)
    \right). \nonumber
\end{eqnarray}
The vacuum is the state annihilated by all the lowering operators: $a_s(\vec p) | \Phi_{Vac} \rangle =0 $ and $b_s(\vec p) | \Phi_{Vac} \rangle =0 $.
The states formed by $(a^\#)^s(\vec p) | \Phi_{Vac}\rangle = |1_{\vec{p},s,+} \rangle $  and 
$(b^\#)^s(\vec p) | \Phi_{Vac}\rangle = |1_{\vec{p},s,-} \rangle $ 
are correctly normalized under the $\eta$ measure.
These ladder operators build the states of the theory.

The operators for the traditional observables are found to agree with the traditional definitions with $\#$ replacing $\dag$.
The Hamiltonian can be re-expressed as
\begin{equation}
    H = \int \frac{d^3\vec p}{(2\pi)^3}  \frac{\omega_p}{2} \,\left( [  (a^\#)^s(\vec p), a_s(\vec p)  ] + [  (b^\#)^s(\vec p), b_s(\vec p)  ] \right).
\end{equation}
The generator of translations from Noether's theorem is
\begin{eqnarray}
    \vec{P} & = & \int d^3\vec{x}\, \pi_\chi^a (\vec x) \vec{\nabla} \chi_a(\vec x) \\
    & = & \int \frac{d^3 \vec{p}}{(2\pi)^3}  \frac{\vec{p}}{2} \,\left( [  (a^\#)^s(\vec p), a_s(\vec p)  ] + [  (b^\#)^s(\vec p), b_s(\vec p)  ] \right). \nonumber 
\end{eqnarray}
For the momentum operator, this definition changes dramatically when we use a finite lattice in the numerical validation, 
which we discuss in detail in Section \ref{Sec1D6PModel}. The possibility that a neighboring lattice site is occupied forces a redefinition of the momentum operator in terms of the more fundamental translation operator.
The angular momentum operator is likewise
\begin{eqnarray}
    J^{jk} & =&  \int d^3\vec{x} \,\pi_\chi^b ( -i x^j   \partial^k + i x^k \partial^j + S^{jk} )_b^{\ a} \chi_a.   
\end{eqnarray}
And the charge operator is  
\begin{eqnarray}
    Q & = &   -i\,\int d^3\vec{x} \, \pi_\chi^a(\vec x)\, (\tilde \gamma^0)_a^{\ b}\, \chi_b(\vec x) \label{EqQSOGeneral3P} \\
    & = & \int \frac{d^3 \vec{p}}{(2\pi)^3}  \frac{1}{2}  \,\left( [  (a^\#)^s(\vec p), a_s(\vec p)  ] - [  (b^\#)^s(\vec p), b_s(\vec p)  ] \right). \nonumber 
\end{eqnarray}
These operators form the set of traditional observables. 

The ladder operators add and subtract energy because they satisfy the expected commutation relations with these observables: 
\begin{eqnarray}
    [ H, a_s(\vec p)] & = & -\omega_p\, a_s(\vec{p}) 
    \\
\,    [ H, (a^\#)^s(\vec{p}) ] & = & \omega_p\, (a^\#)^s(\vec p) 
\\
\,    [ H, b_s(\vec p)] & = & -\omega_p\, b_s(\vec{p}) 
    \\
\,    [ H, (b^\#)^s(\vec{p}) ] & = & \omega_p\, (b^\#)^s(\vec p).
\end{eqnarray}
The ladder operators add or subtract the state's charge as expected because they satisfy: 
\begin{eqnarray}
  \,  [Q, a_s(\vec p)] & = & - a_s(\vec p) \\
 \,   [Q, (a^\#)^s(\vec p)] & = &  (a^\#)^s(\vec p) \\
\,     [Q, b_s(\vec p)] & = &  b_s(\vec p) \\
 \,   [Q, (b^\#)^s(\vec p)] & = & - (b^\#)^s(\vec p) .
\end{eqnarray}


In the next section, we will validate these expressions on a small number of lattice points.

\section{\label{sec:NumericalResults} Numerical Implementation and Results}

The primary purpose of the machinery developed in Sec.~\ref{sec:SecondOrderFermionQuantization} is to provide a discrete model of fermions without the doublers. 
In this section, we will show the results of two numerical discretization studies: 
In \ref{SecModelOnePoint3D} we discuss a stationary fermion mode in $3+1$D modeled as a single point to demonstrate the toolbox, and in \ref{Sec1D6PModel} we show a system of $6$-points in $1+1$D to show the spectrum lacks doublers.

We use large Clifford algebra matrices to represent quantum fields as described by \cite{floreanini1988functional,henneaux1992quantization,surace2022fermionic,friedrich2024holographic}.  The technique is  equivalent to the Jordan-Wigner transformation \cite{Veyrace26050410}.

For $N$ Grassmann degrees of freedom, each consisting of a $\chi$ and $\pi_\chi$ pair, we build a set of $2N$ Clifford matrices $\Gamma_j$ which are $2^N \times 2^N$ and satisfy
\begin{eqnarray}
\,    \{ \Gamma_j, \Gamma_k \} = 2\,\, \unit \, \delta_{j\,k}.
\label{EqDefineGammas}
\end{eqnarray}
The Clifford matrices and associated states are built using a tensor product of Pauli spin matrices such that the $k\,$th qubit from the right is raised by  
\begin{equation}
c^\dag_{k}=\frac{1}{2}(\Gamma_{2k}-i\,\Gamma_{2k+1}),  
\end{equation}
and lowered by its Hermitian conjugate $c_k$. 
The Appendix shows some added detail for these calculations including an explicit representations of $\Gamma$ matrices in Eq.~\ref{EqGammaDef8}.
These are the fundamental \black{(abstract two state, 
$x_i\in\{0,1\}$)} qubits that define spinor space time.
We denote the $2^N$ states with a vector $|\vec x \rangle=| x_{N-1}\, x_{N-2} \ldots x_2 \, x_1\, x_0\, \rangle$ that are built up from the state $| 000 \ldots 000 \rangle$ 
by actions of the field operators $c^\dag_k$ in the order shown by
\begin{eqnarray}
| \vec x \rangle & \equiv & |\, x_{N-1} \, x_{N-2} \, x_{N-3}\, \ldots x_{2}\, x_{1}\, x_{0} \rangle  \nonumber \\
& \equiv & 
(c^\dag_{N-1})^{x_{N-1}} \,   
 \ldots  \, 
(c^\dag_1)^{x_{1}}(c^\dag_{0})^{x_{0}} 
| 0 \rangle.
\label{EqStateDef} 
\label{EqStateDefinitionBuild}
\end{eqnarray}
\black{The rows of the column vector for the wavefunction  are labeled by the qubit states $\{\ket{00\ldots 00},\ket{00\ldots 01}, \ldots,\ket{11\ldots 11}\}$. The zero-indexed row is given by the state label in binary.}

In the discrete case, 
we define the operators $c$ and $c^\dag$ to be unitless.

This is a qubit space.
The  $c_k^\dag$ effectively create excitations in the $k^{th}$  degree of freedom in the spinor qubit-space. 
We have a qubit degree of freedom for every spinor field $\chi_a(x)$ at every spatial point.
The machinery explains how to build space-time states with Lorentz transformation properties out of the qubits.

With this basic notation setup, we proceed to the two examples.

\subsection{A stationary $3+1$D fermion}
\label{SecModelOnePoint3D}

To demonstrate the machinery, we first consider the case for a stationary fermion mode in $d=3$ spatial dimensions.
In this case all the spatial derivatives in the Hamiltonian vanish.
This can therefore be represented as a $3+1$D  quantum field with one point.  
The Majorana representation of the gamma matrices $\tilde \gamma^\mu$ are $N_S \times N_S$, where $N_S=2^{(d+1)/2}=4$ for even $d+1$. We have $4$ fields $\chi_a$ and $4$ fields $\pi_\chi^a$. 
Our state-space includes $N=N_X N_S=4$ qubits; $N_X=1$ for the one spatial point and $N_S=4$  for the four spinors components at each point. 
Superpositions of these four spinor qubit degrees of freedom form the states with definite Lorentz and charge transformation properties for a particle and its antiparticle.
The canonical quantization relations in Eq.~\ref{Eq2ndQuantizationMajoranaConditions} can be satisfied by identifying  
 \black{$8$ total}
Clifford algebra matrices with the $8$ quantum fields as follows:
\begin{eqnarray}
    \chi_j  =  \frac{c_j}{dx}  \ \ \ \ 
    (\pi_\chi)^j  =  \frac{-i\,c^{\dag j}}{dx^2}  \label{EqMajorana1PointOperatorDefs}
\end{eqnarray}
where $j$ goes from $0$ to $3$ and 
where the lattice spacing $dx$ is included to give the fields the canonical units.
These also satisfy the Majorana condition that $\chi=\chi^*$ and $\pi_\chi=-\pi_\chi^*$.
The Hamiltonian for this one stationary mode is 
\begin{eqnarray}
 H_{1}  & = & dx^3 {\mathcal{H}}_{1}  ,     \label{EqH1P} \\
  & = & \frac{(dx)^3}{2} \left( - \pi_\chi^a  (\tilde \gamma^0)_{ac}  \pi_\chi^c   
       +  m^2 \chi_a (\tilde \gamma^0)^{ab} \chi_b \right), \\
    & = &    \frac{1}{dx} c^{\dag\,a}  (\tilde \gamma^0)_{ac}  c^{\dag\,c}   
       +  m^2\,dx\, c_a (\tilde \gamma^0)^{ab} c_b .
       \label{EqH1P}
\end{eqnarray}
Substituting the $c$ and $c^\dag$ matrices into Eq.~\ref{EqH1P} gives a  $16 \times 16$ matrix.
An explicit representation of $H_{1}$ seen in Eq.~\ref{EqH4MajoranaDOFH}.

As seen from Eq.~\ref{EqH4MajoranaDOFH}, the Hamiltonian matrix is not Hermitian.
Even so, $H_1$ still has real eigenvalues which can be shown to be $E_j= \{-2m, -m, -m, -m, -m, 0, 0, 0, 0, 0, 0, m, m, m, m, 2m   \}$.
The pseudo-Hermitian symmetry is
\begin{equation}
    \eta  =  (m\,dx)^{- c_a\,c^{\dag a}}. \label{EqEta1P}
\end{equation} 
As can be seen in Eq.~\ref{EqEta1PNumerical}, the operator puts a factor $({m\,dx})^{-1}$ for each $0$ in the state described in the qubit basis.

The ladder operators are
\begin{eqnarray}
   a_s & = & u^d_s\,\sqrt{\frac{m\,dx}{2}}   \left( c_d + \frac{1}{m\,dx} (\gamma^0)_{db} c^{\dag b} \right),
\end{eqnarray}
\begin{eqnarray}
   b_s & = & v^d_s\,\sqrt{\frac{m\,dx}{2}}   \left( c_d + \frac{1}{m\,dx} (\gamma^0)_{db} c^{\dag b} \right),
\end{eqnarray}
\begin{eqnarray}
   (a^\#)^s & = & (u^\dag)_d^s\,\sqrt{\frac{m\,dx}{2}}   \left( c_a \tilde{\gamma}^{ad} + \frac{1}{m\,dx}  c^{\dag d} \right),
\end{eqnarray}
\begin{eqnarray}
   (b^\#)^s & = & (v^\dag)_d^s\,\sqrt{\frac{m\,dx}{2}}   \left( c_a \tilde{\gamma}^{ad} + \frac{1}{m\,dx}  c^{\dag d} \right).
\end{eqnarray}
These ladder operators define the spin and particle/antiparticle states from the underlying qubits. 
The vacuum state is
\begin{eqnarray}
    |\Phi_{Vac}\rangle & = & \frac{1}{2}\left( 
    (m\,dx)^2 |0000\rangle
    +i (m\,dx) |0110\rangle  \right.  \nonumber \\
  &  & \left. -i (m\,dx) |1001\rangle 
   +|1111\rangle \right)
\end{eqnarray}
and is normalized to satisfy $\langle \Phi_{Vac}|\eta |\Phi_{Vac}\rangle = 1$.

The $15$ excited states are built by acting on the vacuum with the creation operators resulting in the states described in Table \ref{Table16States}. 
In this table the expectation value of the states $| \cdot \rangle$ is taken with the measure $\eta$ as:
$E=\langle \cdot | \eta \, H \,| \cdot \rangle$, 
$J^2=\langle \cdot | \eta \, ( J_x^2 +J_y^2 + J_z^2 )\,| \cdot \rangle = J(J+1)$,  $J_z =\langle \cdot | \eta \, J_z \,| \cdot \rangle$,
$Q=\langle \cdot | \eta \, Q \,| \cdot \rangle$.
This table shows the $16$ states each with a unique set of quantum numbers.
These quantum numbers  correspond exactly with what one expects for a Dirac fermion at a point.

In this example, we have shown how the expected fermion spectrum for stationary modes follows from a pseudo-Hermitian Hamiltonian.
The ladder operators follow  the form one expects from a harmonic oscillator and build the complete set of states.
We have shown how to use the new measure $\eta$ to normalize the states and find expectation values for the observables in terms of these states.
Unlike previous quantization approaches such as FOPV \cite{ferro2024quantization}, this approach does not have excess states and can be equated with a traditional Dirac fermion.

\begin{table*}
\begin{ruledtabular}
\begin{tabular}{cccccl}
State & $E-E_{\mathrm{Vac}}$ & $J^2$ & $J_z$ & $Q$ & State in Spinor Qubit Basis \\
\hline
$a^{\# 0} a^{\# 1} b^{\#0} b^{\#1} |\Phi_{Vac}\rangle$ & $4m$ & $0$ & $0$ & $0$ &
$\frac{1}{2}\left(
 -|1111\rangle + i\,(dx\,m)\,|0110\rangle
 - i\,(dx\,m)\,|1001\rangle
 - (dx\,m)^2\,|0000\rangle
\right)$ \\

$a^{\# 0}  b^{\#0} b^{\#1} |\Phi_{Vac}\rangle$ &
$3m$ & $\frac{3}{4}$ & $\phantom{-}\frac{1}{2}$ & $-1$ &
$\frac{\sqrt{dx\,m}}{2}\left(
 -|1011\rangle + i\,|1101\rangle
 - i\,(dx\,m)\,|0010\rangle
 - (dx\,m)\,|0100\rangle
\right)$ \\

$a^{\# 0}  a^{\#1} b^{\#0} |\Phi_{Vac}\rangle$ &
$3m$ & $\frac{3}{4}$ & $-\frac{1}{2}$ & $\phantom{-}1$ &
$\frac{\sqrt{dx\,m}}{2}\left(
 -|1011\rangle - i\,|1101\rangle
 - i\,(dx\,m)\,|0010\rangle
 + (dx\,m)\,|0100\rangle
\right)$ \\

$a^{\# 1}  b^{\#1} b^{\#0} |\Phi_{Vac}\rangle$ &
$3m$ & $\frac{3}{4}$ & $-\frac{1}{2}$ & $-1$ &
$\frac{\sqrt{dx\,m}}{2}\left(
 -|0111\rangle - i\,|1110\rangle
 - i\,(dx\,m)\,|0001\rangle
 + (dx\,m)\,|1000\rangle
\right)$ \\

$a^{\# 1}  b^{\#1} a^{\#0} |\Phi_{Vac}\rangle$ &
$3m$ & $\frac{3}{4}$ & $\phantom{-}\frac{1}{2}$ & $\phantom{-}1$ &
$\frac{\sqrt{dx\,m}}{2}\left(
 -|0111\rangle + i\,|1110\rangle
 - i\,(dx\,m)\,|0001\rangle
 - (dx\,m)\,|1000\rangle
\right)$ \\

$a^{\# 1}  a^{\#0} |\Phi_{Vac}\rangle$ &
$2m$ & $0$ & $0$ & $\phantom{-}2$ &
$\frac{(dx\,m)}{2}\left(
 |0011\rangle + i\,|0101\rangle
 + i\,|1010\rangle
 - |1100\rangle
\right)$ \\

$b^{\# 1}  a^{\#0} |\Phi_{Vac}\rangle$ &
$2m$ & $2$ & $\phantom{-}1$ & $\phantom{-}0$ &
$\frac{(dx\,m)}{2}\left(
 -|0011\rangle + i\,|0101\rangle
 - i\,|1010\rangle
 - |1100\rangle
\right)$ \\

$b^{\# 1}  b^{\#0} |\Phi_{Vac}\rangle$ &
$2m$ & $0$ & $0$ & $-2$ &
$\frac{(dx\,m)}{2}\left(
 |0011\rangle - i\,|0101\rangle
 - i\,|1010\rangle
 - |1100\rangle
\right)$ \\

$a^{\# 1}  b^{\#0} |\Phi_{Vac}\rangle$ &
$2m$ & $2$ & $-1$ & $\phantom{-}0$ &
$\frac{(dx\,m)}{2}\left(
 -|0011\rangle - i\,|0101\rangle
 + i\,|1010\rangle
 - |1100\rangle
\right)$ \\

$\frac{1}{\sqrt{2}} (a^{\# 1}  b^{\#1}-a^{\# 0}  b^{\#0} ) |\Phi_{Vac}\rangle$ &
$2m$ & $2$ & $\phantom{-}0$ & $\phantom{-}0$ &
$\frac{i\,(dx\,m)}{\sqrt{2}}\left(
 |0110\rangle + |1001\rangle
\right)$ \\

$\frac{1}{\sqrt{2}} (a^{\# 1}  b^{\#1}+a^{\# 0}  b^{\#0} ) |\Phi_{Vac}\rangle$ &
$2m$ & $0$ & $0$ & $\phantom{-}0$ &
$\frac{1}{\sqrt{2}}\left(
 |1111\rangle - (dx\,m)^2\,|0000\rangle
\right)$ \\

$a^{\#1} |\Phi_{Vac}\rangle$ &
$m$ & $\frac{3}{4}$ & $-\frac{1}{2}$ & $\phantom{-}1$ &
$\frac{\sqrt{dx\,m}}{2}\left(
 |1011\rangle + i\,|1101\rangle
 - i\,(dx\,m)\,|0010\rangle
 + (dx\,m)\,|0100\rangle
\right)$ \\

$b^{\#1} |\Phi_{Vac}\rangle$ &
$m$ & $\frac{3}{4}$ & $\phantom{-}\frac{1}{2}$ & $-1$ &
$\frac{\sqrt{dx\,m}}{2}\left(
 -|1011\rangle + i\,|1101\rangle
 + i\,(dx\,m)\,|0010\rangle
 + (dx\,m)\,|0100\rangle
\right)$ \\

$a^{\#0} |\Phi_{Vac}\rangle$ &
$m$ & $\frac{3}{4}$ & $\phantom{-}\frac{1}{2}$ & $\phantom{-}1$ &
$\frac{\sqrt{dx\,m}}{2}\left(
 -|0111\rangle + i\,|1110\rangle
 + i\,(dx\,m)\,|0001\rangle
 + (dx\,m)\,|1000\rangle
\right)$ \\

$b^{\#0} |\Phi_{Vac}\rangle$ &
$m$ & $\frac{3}{4}$ & $-\frac{1}{2}$ & $-1$ &
$\frac{\sqrt{dx\,m}}{2}\left(
 |0111\rangle + i\,|1110\rangle
 - i\,(dx\,m)\,|0001\rangle
 + (dx\,m)\,|1000\rangle
\right)$ \\

$|\Phi_{Vac}\rangle$ &
$0$ & $0$ & $0$ & $\phantom{-}0$ &
$\frac{1}{2}\left(
 |1111\rangle + i\,(dx\,m)\,|0110\rangle
 - i\,(dx\,m)\,|1001\rangle
 + (dx\,m)^2\,|0000\rangle
\right)$ \\
\end{tabular}
\end{ruledtabular}
\caption{
\label{Table16States}
The list of 16 stationary fermion states and their eigenvalues in terms of the spinor qubit basis.
}
\end{table*}

\subsection{A $1+1$D Fermion with $6$ points}
\label{Sec1D6PModel}

Our next example is a model of fermions in $1+1$D with $6$ spatial points on a ring.  
This example requires a modification of the derivative operator such that it commutes with the translation symmetry described in Appendix \ref{SecTorusDerivatives}, but otherwise is a straightforward discritization of the Hamiltonian in Eq.~\ref{EqHamiltonianContinuum}.

\begin{figure*}
    \centering
    (a) \includegraphics[width=0.45\linewidth]{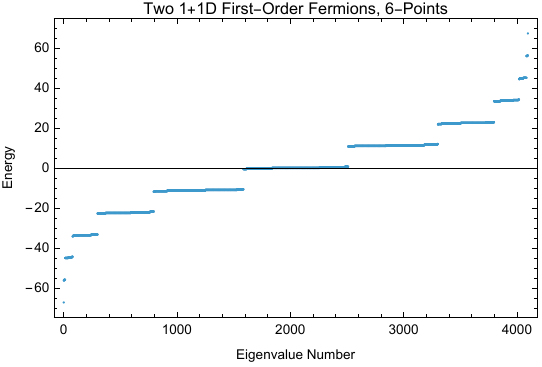}
    (b) \includegraphics[width=0.45\linewidth]{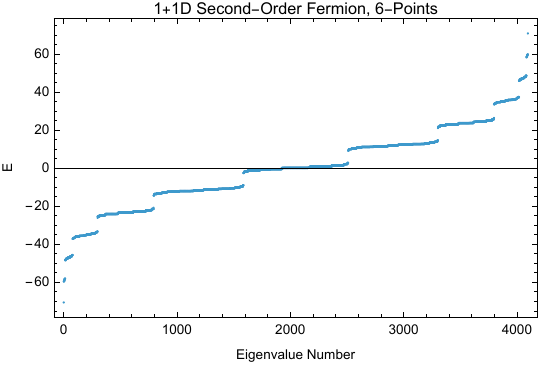}
    \caption{In (a) we show  the 4096 eigenvalues of the a first-order Dirac Hamiltonian with two Majorana fermions at each point.  In (b) we show the 4096 eigenvalues of Hamiltonian in Eq.~\ref{EqH6P}. In both cases, we have taken $m=11$ and $dx=1/3$.  Why does the first-order case look flatter for each band?
    The reason lies in the  dispersion curves in Fig.~\ref{Fig:FirstOrderwSecondOrderDispersion}.  In Fig.~\ref{Fig:FirstOrderwSecondOrderDispersion}(b) one sees $3$ energy values for the momentum states compared to only $2$ energy values in the first order case in Fig.~\ref{Fig:FirstOrderwSecondOrderDispersion}(a).  Therefore the energy spectrum in this figure in the second-order case shows more detail and resolution  achieved with the \black{\textit{same}} sized matrix for $H$.}
    \label{fig:SO6PEnergySpectra}
\end{figure*}

In $1+1D$ the two gamma matrices in the Majorana representation are $\tilde \gamma^0=\sigma_2$ and $\tilde{\gamma}^1=i\sigma_1$.
Our state-space includes $N=N_X N_S=12$ qubits; $N_X=6$ for the six spatial points and $N_S=2$  for the two spinor components at each point. 
The basis for the quantum fields are $24$ $\Gamma_j$ matrices from Eq.~\ref{EqDefineGammas} with each matrix being $4096 \times 4096$.  
In $1+1$D, the field $\chi$ is unitless and field $\pi$ has units of $1/dx$.
The quantum fields for each spinor degree of freedom at each point are given by
\begin{eqnarray}
    \chi_a(j) &= & c_{2j+a} = \frac{1}{2}\left( \Gamma_{2a+4j} + i\,\Gamma_{2a+4j+1} \right) \\
    \pi^a(j) &= & \frac{c^{\dag\,2j+a}}{dx} =  \frac{-i}{2\,dx}\left( \Gamma_{2a+4j} - i\,\Gamma_{2a+4j+1} \right) 
\label{Eq0DSecondOrderFieldQuantization}
\end{eqnarray}
where $j$ and $a$ are \black{zero-based indices}.
With $N_X=6$,  the $1+1$D Hamiltonian is
\begin{eqnarray}
   H_{6} & = &  \sum_{j=0}^{N_X-1} dx  {\mathcal{H}}_j, \\
   & = & \sum_{j=0}^{N_X-1} dx \left(- \pi^a(j) \, (\tilde \gamma^0)_{ab} \,\pi^b(j)  \right.  \nonumber \\
   & & \left. + \frac{1}{2} \chi_a(j) \left( m^2\, \delta^{j,k}  -  (\nabla_T^2)^{j\,k} \right)  \,\chi_b(k)\, (\tilde \gamma^0)^{ab} \right).\qquad \label{EqH6P}
\end{eqnarray}
The operator $\nabla_T^2$  and the translation operator $T_X$ are defined in Appendix \ref{SecTorusDerivatives} to ensure that $[H_{6},T_X]=0$.
We found the eigenvalues of Eq.~\ref{EqH6P} numerically using Mathemtaica assuming $dx=1/3$ and $m=11$ and displayed the results in Fig~\ref{fig:SO6PEnergySpectra} (b).  The $0$, $1$, $2$, $\ldots$ $12$ particle excitations are seen as bands.  
This should be compared to Fig~\ref{fig:SO6PEnergySpectra} (a) with the equivalent first-order Dirac Hamiltonian with two Majorana fermions at each point.  In both cases $H$ is $4096 \times 4096$, but one can see the increased resolution in the second-order-case because the lack of doublers gives more possible energy values for each particle excitation.  
For this reason we claim the second-order approach will be more efficient with computational resources.
The charge operator is 
\begin{equation}
    Q = \sum_{j=0}^{N_X-1} -i\, \pi^A(j)\, (\tilde \gamma^0)_a^{\ b}\, \chi_b(j). \label{EqQSOGeneral6P}
\end{equation}
This operator has eigenvalues of $(-6,-5,-4,-3,-2,-1,0,1,2,3,4,5,6)$.

The $1+1D$ second-order free-fermion discrete case can be solved analytically.
In the next few paragraphs, we first setup the boundary conditions needed to define the mode structure, the ladder operators, and pseudo-Hermitian symmetry $\eta$, and the observables $H$, $Q$, and $P$.

The first challenge is to establish periodic boundary conditions on a lattice. 
We represent our qubits in a particular order and starting point as described in Eq.~\ref{EqStateDef} (or Eq.~\ref{EqStateDefInMultiDims} in higher number of dimensions).
We demand the physics be invariant under a shift of the states by a lattice site.
However, when a derivative operator acts across a boundary, one lowering operator may act on the right-most qubit, while the other lowering operator has to anticommute past all the excited qubits in the state to act on the qubits the opposite left-most side of the state.  
This will pick up a phase of $-1$ if there is an odd number of `1' qubits between the action of the two operators.
The cyclic symmetry we impose therefore manifests itself as acting on two seperate symmetry sectors. 
One finds periodic boundary conditions on the states with even number occupation in the qubit basis and antiperiodic with states with odd number occupation in the qubit basis. 
This is further described in Appendix \ref{SecTorusDerivatives} and is also explained in common fermionic modeling software such as QuSpin \cite[Sec 2.1]{10.21468/SciPostPhys.7.2.020}, in $1$D.

This machinery allows us to describe the Hamiltonian
\begin{eqnarray}
    H_{6}=    H^{(P)}_{6} P_{(P)} + H^{(A)}_{6} P_{(A)}, \label{EqH6PA1}
\end{eqnarray}
where in $H^{(A)}$ the $\nabla^2_T$ is replaced with $\nabla^2_A$ and in $H^{(P)}$ the $\nabla^2_T$ is replaced with $\nabla^2_P$, 
\black{as defined in \ref{delsquaredP} and \ref{delsquaredA}, respectively}.
The index for  periodic and antiperiodic conditions is given by ($\mathcal{B} \in (P,A)$).
The projection operators $P_{(\mathcal{B})}$ and the modes $k^2_{(\mathcal{B})}$ are defined in the appendix in Eq.~\ref{EqProjectors} through Eq.~\ref{EqkSqADef}.
The mode energies are given by 
\begin{equation}
 E_{(\mathcal{B})}(n)=\sqrt{m^2 + k_{(\mathcal{B})}^2(n)}.
\end{equation}

The Fourier modes of both of our fields for the periodic and antiperiodic cases can be written as:
\begin{eqnarray}
    \tilde{\chi}^{(\mathcal{B})}_b(n) & = & U^{(\mathcal{B})\dag}_{nj}\, \chi_b(j), \\
    \tilde{\pi}^{(\mathcal{B})\,b} (n) & = & 
    U^{(\mathcal{B})\dag}_{nj}\, \pi^b(j), \\
    \tilde{\chi}^{*(\mathcal{B})}_b(n) & = &  \chi_b(j) U^{(\mathcal{B})}_{jn}, \\
    \tilde{\pi}^{*(\mathcal{B})b}(n) & = & \pi^{b}(j) U^{(\mathcal{B})}_{jn},
\end{eqnarray}
where
\begin{eqnarray}
U^{(P)}_{jn} & = & \frac{1}{\sqrt{N_X}} e^{+i\,\frac{j\,n\,2\pi}{N_X}},
\\
U^{(A)}_{jn} & = &\frac{1}{\sqrt{N_X}} e^{+i\,\frac{j\,(n+\frac{1}{2})\,2\pi}{N_X}} .
\end{eqnarray}
We have chosen to enumerate the indices that run through the momentum space modes $1 \ldots N_X$.

The eigenvectors of $\tilde \gamma^0$ are 
\begin{equation}
    u_p = \frac{1}{\sqrt{2}}( 1, i) \ \ \ \
    v_n = \frac{1}{\sqrt{2}}( 1, -i) 
\end{equation}
such that $\tilde \gamma^0 \, u_p = +u_p$ and $\tilde \gamma^0 \, v_n = -v_n$.
There is only one spin-state for each charge so there are no indices on $u$ or $v$.  We use these to form the  annihilation operators for positive and negative charge as 
\begin{eqnarray}
\hspace{-0.25in}
    a^{(\mathcal{B})}(n) & =& (u_p)^c \frac{\sqrt{E_{(\mathcal{B})}(n)\,dx}}{\sqrt{2}} \left( \tilde{\chi}^{(P)}_c(n) + i \frac{(\tilde{\gamma}^0)_{cb}}{E_{(\mathcal{B})}(n)}\,  \tilde{\pi}^{(\mathcal{B})\,b}(n) \right) \nonumber \\
\hspace{-0.25in}
    b^{(\mathcal{B})}(n) & = & (v_n)^c \frac{\sqrt{E_{(\mathcal{B})}(n)\,dx}}{\sqrt{2}} \left( \tilde{\chi}^{(P)}_c(n) + i \frac{(\tilde{\gamma}^0)_{cb}}{E_{(\mathcal{B})}(n)}\,  \tilde{\pi}^{(\mathcal{B})\,b}(n) \right). \nonumber
\end{eqnarray}
The raising operators are given by
\begin{eqnarray}
     (a^{\#})^{(\mathcal{B})}=(a^{(\mathcal{B})})^{\#} \equiv \eta^{-1} \,(a^{(\mathcal{B})})^\dag \, \eta \\
      (b^{\#})^{(\mathcal{B})}=(b^{(\mathcal{B})})^{\#} \equiv \eta^{-1} \,(b^{(\mathcal{B})})^\dag \, \eta.
\end{eqnarray}

The psuedo-Hermitian measure for the periodic and antiperiodic subspace is
\begin{equation}
\hspace{-0.35in}
    \eta^{(\mathcal{B})}  =  \exp\left( -i\, \sum_{n=1}^{N_X} \log(\,E_{(\mathcal{B})}(n)\, dx\,)\ \tilde{\xi}^{*(\mathcal{B})}_c(n) \ \tilde{\pi}^{(\mathcal{B})c}(n)  \right)    
\end{equation}
and the pseudo-Hermitian measure for $H_6$ is then
\begin{equation}
    \eta = P_{(A)} \eta^{(A)} + P_{(P)} \eta^{(P)}.
\end{equation}
The periodic and antiperiodic Hamiltonian in Eq.~\ref{EqH6PA1} can equivalently be expressed as 
\begin{eqnarray}
    H^{(\mathcal{B})}_6  & = &   \sum_{n=1}^{N_X} \frac{E_{(\mathcal{B})}(n)}{2} \left[  (a^\#)^{(\mathcal{B})}(n), a^{(\mathcal{B})}(n) \right]
    \nonumber \\
     & & +
    \frac{E_{(\mathcal{B})}(n)}{2} \left[  (b^\#)^{(\mathcal{B})}(n), b^{(\mathcal{B})}(n) \right].
    \label{EqH6ModeDecomposition}
\end{eqnarray}

To form the complete set of states we must identify the minimum energy state of $H^{(P)}$ and $H^{(A)}$. 
From Eq.~\ref{EqH6ModeDecomposition} and the definition that $a^{(\mathcal{B})}(n)$ and $b^{(\mathcal{B})}(n)$  give $0$ when acting on the vacuum
$|\Phi^{(\mathcal{B})}_{V} \rangle$, we can deduce the minimum energy state energy corresponds to 
\begin{equation}
    E^{(\mathcal{B})}_{Min}=- 2 \sum_{n=1}^{N_X} \frac{E_{(\mathcal{B})}(n)}{2}
\label{EqMinEnergy1DMajoranaRealSecondOrder}
\end{equation}
where the $1/2 $ is for each mode and the $2$ is because there are modes for particles and antiparticles.
They satisfy
\begin{equation}
    H_6^{(\mathcal{B})} |\Phi^{(\mathcal{B})}_V \rangle =E^{(\mathcal{B})}_{Min}\,  |\Phi^{(\mathcal{B})}_V \rangle  
\end{equation}
where there is no sum on $\mathcal{B}$. 
The states are then built by the raising and lower operators with the respective boundary conditions.
Because $E_{Min}^{(A)} < E_{Min}^{(P)}$, the true vacuum is the antisymmetric minimum energy state $|\Phi_V^{(A)} \rangle$.
This matches the numerical result ($E_{Vac} = E^{(A)}_{Min}=-70.6645$) when we substitute $m=11$ and $dx=1/3$ into Eq~\ref{EqH6P} and calculate the minimum eigenvalue.

The set of excited states for one-particle corresponds to $a^{\#(P)}(n) |\Phi^{(P)}_{V}\rangle$.
The set of excited states of two-particles corresponds to $a^{\#(A)}(n)\, a^{\#(A)}(n') |\Phi^{(A)}_{V}\rangle$. 
Let the set of excited states of three-particles corresponds to $a^{\#(P)}(n)\, a^{\#(P)}(n')\,a^{\#(P)}(n'') |\Phi^{(P)}_{V}\rangle$. 
Negatively charged states can be built from $b^\#$ operators.
\black{The variables $p_j$ and $n_j$ represent the number of excitations (either $0$ or $1$) for each positive charge or negative negative charge with momentum mode $j$.}  
We define the excited states with the phase such that
\begin{eqnarray}
 | n( n_6 \ldots n_1 ) p( p_6 \ldots p_1) \rangle   \hfill\nonumber 
 \\
\equiv (b^{\#(\mathcal{B})}(6))^{n_6} \ldots (a^{\#(\mathcal{B})}(1))^{p_1} |\Phi^{(\mathcal{B})}_V\rangle   
\end{eqnarray}
where $\mathcal{B}$ is $P$ if $\sum_j n_j+p_j$ is odd and $\mathcal{B}$ is $A$ otherwise.   
We defined the $\eta$, vacuum state, and the creation operators such that the resulting states are normalized:
\begin{eqnarray}
    \langle n( n'_6 \ldots n'_1 ) p( p'_6 \ldots p'_1) | \,\eta\,  | n( n_6 \ldots n_1 ) p( p_6 \ldots p_1) \rangle \nonumber \\
   =  
  \delta_{n_1}^{n'_1}  \ldots \delta_{n_6}^{n'_6}
     \delta_{p_1}^{p'_1}  \ldots 
     \delta_{p_6}^{p'_6}.\qquad \label{EqMomenta2D6Points}
\end{eqnarray}
These are also eigenstates of the charge and momentum operators. Under the charge operator $Q$ :
\begin{eqnarray}  
    \langle n( n'_6 \ldots n'_1 ) p( p'_6 \ldots p'_1) | \,\eta\,  Q\,| n( n_6 \ldots n_1 ) p( p_6 \ldots p_1) \rangle \nonumber \\
    =
    \sum_{j=1}^{N_X} p_j-n_j.\qquad
\end{eqnarray}
The momentum operator can be extracted from the translation operator defined in Appendix \ref{SecTorusDerivatives}.   Translation by one lattice site indices the following phase:
\begin{widetext} 
\begin{eqnarray*}
    \langle n( n'_6 \ldots n'_1 ) p( p'_6 \ldots p'_1) | \,\eta\, T_X\, | n( n_6 \ldots n_1 ) p( p_6 \ldots p_1) \rangle \equiv \exp(-i \phi) \times  \delta_{n_6}^{n'_6}  \ldots \delta_{n_1}^{n'_1}
     \delta_{p_6}^{p'_6}  \ldots 
     \delta_{p_1}^{p'_1}     \nonumber \\
     = \exp\left( -i\,\frac{2\pi}{N_X}
    \sum_{j=1}^{N_X} \left(j+
    \frac{
\sum_{k=1}^{N_X} n_k+p_k+1 \Mod 2}{2} \right) (n_j+p_j) \right)   
    \times  \delta_{n_6}^{n'_6}  \ldots \delta_{n_1}^{n'_1}
     \delta_{p_6}^{p'_6}  \ldots 
     \delta_{p_1}^{p'_1}. \label{EqMomenta2D6Points}
\end{eqnarray*}
\end{widetext}
The momentum eigenvalue is the phase $\phi$ that results from this translation of the state by one lattice site divided by the lattice spacing $dx$.
The term $\frac{
\sum_{k=1}^{N_X} n_k+p_k+1 \Mod 2}{2}$ gives $1/2$ for states with even number of excitations and $0$ otherwise.
When the phase is greater than $\pi$,  this is a negative momentum state equivalent to \black{ $-(2\pi-\phi)/dx$}. 

There are a total of $4096$ states that can be built from the vacuum and six positive \black{charge} creation operators $a^\#$ and six negative \black{charge} creation operators $b^\#$.
To make sense of these states, we will look a few subsets.

First we study the dispersion relation. 
In Fig.~\ref{Fig:FirstOrderwSecondOrderDispersion} (b), 
we show the momentum and energy eigenvalues \black{(red dots)} of states with a single excitation.
The resulting spectrum makes  clear that there are no doublers which demonstrates the main claim of the paper.

\begin{figure}
    \centering
    \includegraphics[width=1\linewidth]{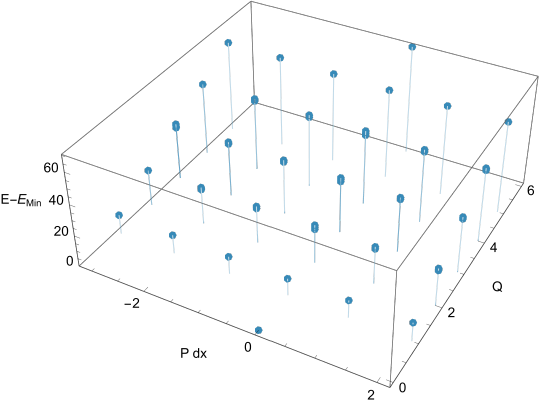}
    \caption{Showing positive $1+1$D states from a second-order fermion model with $6$ points on the ring where $dx=1/3$ and $m=11$.  
    The $64$ states built with only positive ladder operators. The vacuum is shown as the $P=Q=E-E_{Min}=0$ state.
    The charges range from $0$ to $6$ and the $6$ momentum eigenvalues of $P\,dx$ range from $-\pi$ to $\pi$. \black{Stems are added below the states to better show the 3D nature of the $P$, $Q$, $E$ relationships.}
    }
    \label{Fig:6PQPEPlot}
\end{figure}

To further illustrate the states, 
we plotted the $64$  positive excitations vs charge, energy, and momentum in Fig.~\ref{Fig:6PQPEPlot} where we again set $dx=1/3$ and $m=11$. 
The vacuum is shown as the $P=Q=E-E_{Min}=0$ state.
The charges range from $0$ to $6$ and the $6$ momentum eigenvalues of $P\,dx$ range from $-\pi$ to $\pi$. 
When there are $6$ positive excitations, all the momentum states are filled and the total momentum is $P=0$.

\section{\label{sec:Discussion} Discussion and Conclusions}

\begin{table*}
\hspace{-4.5em}
\begin{tabular}{cccc}\hline\hline
Quantity & First Order Dirac & Second-Order Fermion & Notes \\
\hline
Lagrangian Density (${\mathcal{L}}$) & $\psi^\dag \gamma^0 (\gamma^\mu \partial_\mu - m) \psi$ & $   
    \frac{-1}{2} \partial_\mu \chi\ \tilde \gamma^0 \partial^\mu \chi   - \frac{1}{2} m^2 \chi \tilde \gamma^0 \chi$ & Both Lorentz invariant \\
Conjugate Momenta & $\pi^a_\psi = -i\,(\psi_a)^\dag$ & $\pi^a_\chi = (\tilde \gamma^0)^{ab} \partial^0 \chi_b$ & Second order $\pi_\chi$ can be $\pm$ \\
Fields and Units in $3+1$D & $[ \psi ] = dx^{-3/2} $ & $[ \chi ] = dx^{-1}$, $[\pi_\chi] = dx^{-2}$ & $\psi$ complex. $\chi$ is real.  \\
Degrees of Freedom & 4 Complex Fields ($\psi$ and $\psi^\dag$) & 8 Real Fields ($\chi$ and $\partial_0 \chi$) & Equal numbers of DoF \\
Complex conjugation & $\psi=\xi_R + i\xi_I$, $\psi^* = \xi^*_R - i\xi^*_I$ & $\chi^* = \chi$ and $\pi^* = - \pi$ &   \hspace{-0.25in} Real Grassmann  variables: $(ab)^*=ba$  \\ 
Hermitian conjugation & $\psi=\xi_R + i\xi_I$, $\psi^\dag = \xi_R - i\xi_I$ &  $\chi^\dag = \frac{\delta}{\delta \chi}$ &  HC follows Ref.~\cite{floreanini1988functional} \\
Quantization & $\{\psi_a(x), \psi_b^\dag(x') \} = \delta_{ab} \delta(x-x')$ & $\{\chi_b(x),\pi^a_\chi(x') \} = -i \delta^a_b \delta(x-x') $ & Canonical Quantization \\
Hamiltonian Density  (${\mathcal{H}}$)  & 
$ \psi^\dag  \gamma^0 \gamma^j (-i\partial_j)\psi + m \psi^\dag \gamma^0 \psi$ & $
     \frac{-1}{2} (\pi_\chi)  (\tilde \gamma^0)  (\pi_\chi)   
      - \frac{1}{2}  \chi (\tilde \gamma^0) ( \nabla^2 -m^2) \chi $  &  Both have real eigenvalues. \\
Hamiltonian ($H = \int d^3x {\mathcal{H}}$) & $H=H^\dag$ & $H^\# = \eta^{-1} H^\dag \eta$ = H & Second-order is Pseudo-Hermitian \\
Global $U(1)$ Symmetry & $\psi \rightarrow \exp(i\theta) \psi$ & $\chi \rightarrow \exp(i\,\theta \tilde \gamma^0) \chi$ & Distinct Complex Phase \\  
Charge ($Q$) & $\int d^3x \psi^\dag \psi > 0 $  & $-i\,\int d^3\vec{x} \, \pi_\chi \tilde{\gamma}^0 \chi = -i\,\int d^3x \chi \partial_t \chi$ & Both signs possible in $2$nd order \\
Number Operator &  $a^\dag_{\vec{p},s} a_{\vec{p},s}$ &  $(a^{\# s})(\vec p) a_s(\vec{p})$   &  Note $\#$ replaces $\dag$ \\  
State Normalization & $\langle \Phi | \Phi \rangle = 1$ &   $\langle \Phi |\eta | \Phi \rangle = 1$ & Second-Order inner products use $\eta$ \\
Lorentz-Invariant Scalar & $\psi^\dag \gamma^0 \psi$ & $\chi_a (\tilde{\gamma}^0)^{ab}  \chi_b$ & Second-Order uses Majorana Rep. \\
Lattice Translation Operator & $T_X$  & $T_X$  & Same on a Lattice \\
Lattice Momentum Operator & $(-i/dx) \log T_X$ & $(-i/dx) \log T_X$ & Same on a Lattice\\
\hline\hline
\end{tabular}
\caption{\label{TableFOSOComparison} Comparison between first-order Dirac and second-order fermion quantities.}
\end{table*}

Dirac built 
\black{his} equation
to find a relativistic first-order
wave-equation for quantum mechanics.
After unpacking Dirac's discovery, history shows we also learned that the Clifford algebra generates representations of the Lorentz algebra in which the Dirac equation is a special case. 
We also learned that the Dirac field should be treated as an anticommuting quantum field so as to have Fermi-Dirac statistics.
In this paper we retain the transformation properties and the anticommuting nature of the quantum fields, but we return to a second-order formulism.

The second-order fermion description provides a framework that is both relativistic, fermionic, and reclaims the intuition offered by a canonical formulism with coordinates and momenta.  In Table \ref{TableFOSOComparison}, we compare the first order and second order descriptions of fermions.

The second-order technique has a number of benefits.  
The first benefit is the lack of doublers in the lattice formulation as seen in the comparison depicted in Fig.~\ref{Fig:FirstOrderwSecondOrderDispersion} between the first-order case in (a) and the second-order case in (b). The lack of doublers comes about because the numerical implementation of a second-order derivative does not skip a lattice site and therefore does not introduce extra aliasing. 

Also, by quantizing the real Grassmann representation (as opposed to the complex representation) and identifying the $U(1)$ symmetry in this real representation, we have avoided the extra states required to address the quantization condition $\{ \chi^\dag,\chi \} = 0 $ present in previous second-order fermion quantization approaches such as FOPV \cite{ferro2024quantization}. 
Therefore, the states described by our second-order formulism can now be equated with a traditional Dirac fermion.

Next, the fields and their conjugate momenta again have clear roles.  The $\chi$ are unambiguously coordinates where $\pi_\chi$ clearly represents the conjugate momenta.
In the first-order description, the coordinates and the their momenta are mixed \cite{feshbach1958elementary}.

Also, the second-order description improves use of limited computational resources.
Because there are no doublers, fewer qubits are needed for better results, and the system smoothly approaches the continuum limit. 
See for example the improved resolution on energy spectra in Fig.~\ref{fig:SO6PEnergySpectra} where both first and second order free fermion systems are solved with the same size Hamiltonian $(4096 \times 4096)$.
The reason lies in the  dispersion curves in Fig.~\ref{Fig:FirstOrderwSecondOrderDispersion}.  In Fig.~\ref{Fig:FirstOrderwSecondOrderDispersion} (b) one sees $3$ energy values for the momentum states compared to only $2$ energy values in the first order case in Fig.~\ref{Fig:FirstOrderwSecondOrderDispersion} (a).  Therefore the second-order fermion energy spectrum in Fig.~\ref{fig:SO6PEnergySpectra} (b) shows more detail and resolution  achieved than in (a) with the same sized matrix for $H$.

\black{As described by Davoudi \cite{davoudi2025tasi}, some remedies to the fermion doubling problem work in $1D$, but leave doublers when applied to  higher dimensions such as $3+1$D.
The choices for which doublers to suppress are referred to as `tastes'.
When modeling fermions in $3+1$D using our technique, there are no `tastes'.} 
The techniques described in the Appendix ensure the second derivatives can be defined in arbitrary number of spatial dimensions. 

Furthermore, there is nothing in the second-order technique we described that prohibits the $m=0$ limit.
For example, the Hamiltonian Eq.~\ref{EqH6P}, the charge $Q$ in Eq.~\ref{EqQSOGeneral6P}, and the momentum in Eq.~\ref{EqMomenta2D6Points} do not suffer any ill behavior when $m=0$.
One might notice that $\eta$ in the case of one point in Eq.~\ref{EqEta1P} (and written out explicitly in Eq.~\ref{EqEta1PNumerical}) has a division by $m$.  However, this is really a division by the energy \black{$\omega_q$} of the excited mode as seen in Eq.~\ref{EqEtaDef}.
In the case of a single point the only excited mode has energy $m$.  
In the massless case, every excited mode has a non-zero energy and therefore there is no obstruction.

The new perspective has potential to provide insight to long-standing questions.  
The coordinate and conjugate momenta perspective shows opposite charges manifest as coordinates performing opposite rotations on an internal space.  This can be seen because positive and negative \black{charge} states exchange by $\pi \rightarrow -\pi$.
The model shows a close tie between energy conservation and probability conservation suggesting some potential interesting insights into the foundations of quantum mechanics.
Does the new perspective change our view of the nature of vacuum energy?  
Does the new perspective change our view of the nature of the mass-gap problem?  
Both questions can now be more clearly interpreted, suggesting they would be good to revisit with this new perspective.

There is much work left to be done.
Our work does not yet solve the problem of modeling the electroweak theory on a lattice.
We have yet to demonstrate how the model works with a local gauge theory.
We currently foresee no  obstructions to representing the gauge fields on the links between lattice sites as was done by Kogut and Susskind \cite{Kogut.PhysRevD.11.395}. 
The measure $\eta$ depends on the mode decomposition, so there may  need to be some adaptation when the system cannot be analytically solved.
Also, it is not clear if one can represent a left-handed state without a right-handed state.
We performed the modeling here in Mathematica, but it would be valuabe to  translate the model to common codes such as QuSpin \cite{10.21468/SciPostPhys.2.1.003}.
Last it would be valuable to fully estimate the quantum resources this technique may save if implemented in a quantum computer \cite{davoudi2025tasi}.

In conclusion, we have shown  three novel features unique to our second-order fermion technique:  
First we identify the global $U(1)$ symmetry present in a second-order real Majorana fermion model that corresponds to the $U(1)$ \black{symmetry} in the first-order formulation.
As such, the fermions described can be identified with an ordinary electron or positron.
Second, we identified a new pseudo-Hermitian symmetry $\eta$ specific to this model.
Last, we demonstrate that when quantized on a lattice the fermion doublers disappear in this system.  
The described technique works in higher number of dimensions and in the $m=0$ limit.

\begin{acknowledgments}

We are grateful to Niklas Muller, Lucas Kovalsky, Andy Landahl, David Meyer, and Ivan Avramidi for helpful conversations. 
MS also thanks Elvis, D, Matt, JC, Angel, Sameer, Dan, and Al for their support enabling participation in the DoD SkillBridge program at NMT.
MS also thanks  Sharon Sessions, Raúl Morales Juberias, Mike Jackson, Ivan Deutsch, and Craig Keif for hosting me during this work at NMT, UNM CQuIC and UNM COSMIAC research centers  as a temporary visitor during this work.
Most importantly, MS thanks Laura Serna for edits and encouragement.
\smallskip

\black{The Mathematica codes used to produce this data are available upon request.}
\end{acknowledgments}

\appendix


\section{Additional Details for Numerical Implementation}
\label{SecAppendixCalcDetailsOnePoint}

In the appendices, we provide some additional details for the calculations described in the body of the paper.

First we clarify how we construct the Clifford matrices used to represent fermionic quantum fields.
A stationary fermionic degree of freedom (e.g. an electron or positron) at a single point, as described in Sec.~\ref{SecModelOnePoint3D}, requires the index $i$ of the Clifford algebra to run from $0$ to $7$ which will act on a space of dimension $2^{8/2}=16$.
We can construct this set of Clifford matrices $\Gamma_j$  using a tensor product  of Pauli-spin matrices
\black{\cite{surace2022fermionic,Veyrace26050410}:}
\begin{eqnarray}
  \Gamma_{0} =  
                           \sigma_3 \otimes  \sigma_3 \otimes  \sigma_3 \otimes \sigma_1  \nonumber \\
  \Gamma_{1} = 
                           \sigma_3 \otimes  \sigma_3 \otimes  \sigma_3 \otimes \sigma_2  \nonumber \\
  \Gamma_{2} = 
                           \sigma_3 \otimes  \sigma_3 \otimes  \sigma_1 \otimes \sigma_0  \nonumber \\
  \Gamma_{3} = 
                           \sigma_3 \otimes  \sigma_3 \otimes  \sigma_2 \otimes \sigma_0  \nonumber \\
  \Gamma_{4} = 
                           \sigma_3 \otimes  \sigma_1 \otimes  \sigma_0 \otimes \sigma_0  \nonumber \\
  \Gamma_{5} =  
                           \sigma_3 \otimes  \sigma_2 \otimes  \sigma_0 \otimes \sigma_0  \nonumber \\ 
  \Gamma_{6} =  
                           \sigma_1 \otimes  \sigma_0 \otimes  \sigma_0 \otimes \sigma_0  \nonumber \\
  \Gamma_{7} =                            \sigma_2 \otimes  \sigma_0 \otimes  \sigma_0 \otimes \sigma_0.  
  \label{EqGammaDef8}
 \end{eqnarray}
The tensor product is defined in terms of:
\begin{equation}
A = \left(
\begin{array}{cc}
 A_{00} & A_{01} \\
 A_{10} & A_{11} \\
\end{array}
\right) \ \ \  B = \left(
\begin{array}{cc}
 B_{00} & B_{01} \\
 B_{10} & B_{11} \\
\end{array}
\right)
\end{equation}
where
\begin{equation}
\vspace{.2cm}
A \otimes B = \left(
\begin{array}{cccc}
 A_{00} B_{00} &
   A_{00} B_{01} &
    A_{01} B_{00} &
    A_{01} B_{01} \\
   A_{00} B_{10} &
   A_{00} B_{11} &
    A_{01} B_{10} &
    A_{01} B_{11} \\ 
  A_{10} B_{00} &
   A_{10} B_{01} &
    A_{11} B_{00} &
    A_{11} B_{01} \\
   A_{10} B_{10} &
   A_{10} B_{11} &
    A_{11} B_{10} &
    A_{11} B_{11} \\
\end{array}
\right).
\end{equation} 
Our Pauli matrices are:
 \begin{eqnarray}
   (\sigma_0) = \left(
\begin{array}{cc}
 1 & 0 \\
 0 & 1 \\
\end{array}
\right)    
\ \ \ 
   (\sigma_1) = \left(
\begin{array}{cc}
 0 & 1 \\
 1 & 0 \\
\end{array}
\right)
\ \ \   \\
   (\sigma_2) = \left(
\begin{array}{cc}
 0 & -i \\
 i & 0 \\
\end{array}
\right)
\ \ \
   (\sigma_3) = \left(
\begin{array}{cc}
 1 & 0 \\
 0 & -1 \\
\end{array}
\right).
 \end{eqnarray}
Larger families of Clifford matrices can be built by including more tensor products following the above pattern.

 There is an equivalent alternative manner to build arbitrarily large Clifford algebras through recursion.
This is the method used by Friedrich, Cao, and Carroll \cite{friedrich2024holographic}.
Here is the formula producing the $2 \,N$ matrices of the Clifford Algebra $Cl_{(2\,N)}(j) \equiv \Gamma_j$ for $j$ between $0$ and $2N-1$:
\begin{equation}
  Cl_{(2\,N)} (j) = 
\left\{
  \begin{array}{l }
 {\rm{if}}\ N=1 \ {\rm{return\ }}       \\
 \hspace{1cm} \sigma_{j+1}  \\
 {\rm{if}}\ N>1 \ {\rm{return\ }}    \\
\hspace{1cm} \sigma_3 \otimes Cl_{(2N-2)}(j) \ \ {\rm{for}}\  j<2n   \\
 \hspace{1cm} \sigma_1 \otimes \unit_{(2N-2)}  \ \ {\rm{for}}\  j=2n  \\
 \hspace{1cm} \sigma_2  \otimes \unit_{(2N-2)}   \ \ {\rm{for}}\  j=2n+1   .
\end{array} \right.
\end{equation}
Each matrix $\Gamma_j$ is $2^N \times 2^N$.

Once the quantum fields are represented as $\Gamma_j$ matrices and substituted into the Hamiltonian in Eq.~\ref{EqH1P}, we arrive at the pseudo-Hermitian matrix which represents $H_{1}$:
\begin{widetext}
\begin{equation}
\tiny
    H_{1} =
    \left(
\begin{array}{cccccccccccccccc}
 0 & 0 & 0 & 0 & 0 & 0 & -i \text{dx} m^2 & 0 & 0 & i \text{dx} m^2 & 0 & 0 & 0 & 0 & 0 & 0 \\
 0 & 0 & 0 & 0 & 0 & 0 & 0 & -i \text{dx} m^2 & 0 & 0 & 0 & 0 & 0 & 0 & 0 & 0 \\
 0 & 0 & 0 & 0 & 0 & 0 & 0 & 0 & 0 & 0 & 0 & -i \text{dx} m^2 & 0 & 0 & 0 & 0 \\
 0 & 0 & 0 & 0 & 0 & 0 & 0 & 0 & 0 & 0 & 0 & 0 & 0 & 0 & 0 & 0 \\
 0 & 0 & 0 & 0 & 0 & 0 & 0 & 0 & 0 & 0 & 0 & 0 & 0 & -i \text{dx} m^2 & 0 & 0 \\
 0 & 0 & 0 & 0 & 0 & 0 & 0 & 0 & 0 & 0 & 0 & 0 & 0 & 0 & 0 & 0 \\
 \frac{i}{\text{dx}} & 0 & 0 & 0 & 0 & 0 & 0 & 0 & 0 & 0 & 0 & 0 & 0 & 0 & 0 & i \text{dx} m^2
   \\
 0 & \frac{i}{\text{dx}} & 0 & 0 & 0 & 0 & 0 & 0 & 0 & 0 & 0 & 0 & 0 & 0 & 0 & 0 \\
 0 & 0 & 0 & 0 & 0 & 0 & 0 & 0 & 0 & 0 & 0 & 0 & 0 & 0 & -i \text{dx} m^2 & 0 \\
 -\frac{i}{\text{dx}} & 0 & 0 & 0 & 0 & 0 & 0 & 0 & 0 & 0 & 0 & 0 & 0 & 0 & 0 & -i \text{dx} m^2
   \\
 0 & 0 & 0 & 0 & 0 & 0 & 0 & 0 & 0 & 0 & 0 & 0 & 0 & 0 & 0 & 0 \\
 0 & 0 & \frac{i}{\text{dx}} & 0 & 0 & 0 & 0 & 0 & 0 & 0 & 0 & 0 & 0 & 0 & 0 & 0 \\
 0 & 0 & 0 & 0 & 0 & 0 & 0 & 0 & 0 & 0 & 0 & 0 & 0 & 0 & 0 & 0 \\
 0 & 0 & 0 & 0 & \frac{i}{\text{dx}} & 0 & 0 & 0 & 0 & 0 & 0 & 0 & 0 & 0 & 0 & 0 \\
 0 & 0 & 0 & 0 & 0 & 0 & 0 & 0 & \frac{i}{\text{dx}} & 0 & 0 & 0 & 0 & 0 & 0 & 0 \\
 0 & 0 & 0 & 0 & 0 & 0 & -\frac{i}{\text{dx}} & 0 & 0 & \frac{i}{\text{dx}} & 0 & 0 & 0 & 0 & 0
   & 0 \\
\end{array}
\right).
\label{EqH4MajoranaDOFH}
\end{equation}
\normalsize
\black{The rows and columns of $H_{1}$ are labeled by the qubit states $\{\ket{0000},\ket{0001}, \ldots,\ket{1111}\}$.}
From the matrix representation, it is clear that $H_{1}$ is not Hermitian.
The pseudo-Hermitian symmetry from Eq.~\ref{EqEta1P} is given explicilty by
\tiny
\begin{equation}
    \eta_{1P}=\left(
\begin{array}{cccccccccccccccc}
  \frac{1}{\text{dx}^4 m^4} & 0 & 0 & 0 & 0 & 0 & 0 & 0 & 0 & 0
   & 0 & 0 & 0 & 0 & 0 & 0 \\
 0 & \frac{1}{\text{dx}^3 m^3} & 0 & 0 & 0 & 0 & 0 & 0 & 0 & 0
   & 0 & 0 & 0 & 0 & 0 & 0 \\
 0 & 0 & \frac{1}{\text{dx}^3 m^3} & 0 & 0 & 0 & 0 & 0 & 0 & 0
   & 0 & 0 & 0 & 0 & 0 & 0 \\
 0 & 0 & 0 & \frac{1}{\text{dx}^2 m^2} & 0 & 0 & 0 & 0 & 0 & 0
   & 0 & 0 & 0 & 0 & 0 & 0 \\
 0 & 0 & 0 & 0 & \frac{1}{\text{dx}^3 m^3} & 0 & 0 & 0 & 0 & 0
   & 0 & 0 & 0 & 0 & 0 & 0 \\
 0 & 0 & 0 & 0 & 0 & \frac{1}{\text{dx}^2 m^2} & 0 & 0 & 0 & 0
   & 0 & 0 & 0 & 0 & 0 & 0 \\
 0 & 0 & 0 & 0 & 0 & 0 & \frac{1}{\text{dx}^2 m^2} & 0 & 0 & 0
   & 0 & 0 & 0 & 0 & 0 & 0 \\
 0 & 0 & 0 & 0 & 0 & 0 & 0 & \frac{1}{\text{dx} m} & 0 & 0 & 0
   & 0 & 0 & 0 & 0 & 0 \\
 0 & 0 & 0 & 0 & 0 & 0 & 0 & 0 & \frac{1}{\text{dx}^3 m^3} & 0
   & 0 & 0 & 0 & 0 & 0 & 0 \\
 0 & 0 & 0 & 0 & 0 & 0 & 0 & 0 & 0 & \frac{1}{\text{dx}^2 m^2}
   & 0 & 0 & 0 & 0 & 0 & 0 \\
 0 & 0 & 0 & 0 & 0 & 0 & 0 & 0 & 0 & 0 & \frac{1}{\text{dx}^2
   m^2} & 0 & 0 & 0 & 0 & 0 \\
 0 & 0 & 0 & 0 & 0 & 0 & 0 & 0 & 0 & 0 & 0 &
   \frac{1}{\text{dx} m} & 0 & 0 & 0 & 0 \\
 0 & 0 & 0 & 0 & 0 & 0 & 0 & 0 & 0 & 0 & 0 & 0 &
   \frac{1}{\text{dx}^2 m^2} & 0 & 0 & 0 \\
 0 & 0 & 0 & 0 & 0 & 0 & 0 & 0 & 0 & 0 & 0 & 0 & 0 &
   \frac{1}{\text{dx} m} & 0 & 0 \\
 0 & 0 & 0 & 0 & 0 & 0 & 0 & 0 & 0 & 0 & 0 & 0 & 0 & 0 &
   \frac{1}{\text{dx} m} & 0 \\
 0 & 0 & 0 & 0 & 0 & 0 & 0 & 0 & 0 & 0 & 0 & 0 & 0 & 0 & 0 & 1
   \\
\end{array}
\right)
\label{EqEta1PNumerical}
\end{equation}
\normalsize
\end{widetext}

The symmetry $H^\# \equiv\eta^{-1} H^\dag \eta = H$ can now be explicitly verified.
These matrices also alow one to explicitly check that states from Table \ref{Table16States} are normalized and those states are eigenvectors of $H_1$.

These additional detail should provide the reader the ability to build an understanding of the toolbox presented in a simple example.

\section{Solving for $\eta$}
\label{SecEtaSolution}

To find the solution to Eq.~\ref{EqPiSymmetry} and \ref{EqChiSymmetry}, we use 
\black{the functional identity:}
\begin{eqnarray}
 \frac{\delta }{\delta c_a(p)} \exp \left( \int dq\, g(q) c_b(q) \frac{\delta }{\delta c_b(q)}  \right)  = \nonumber  \\
 \exp( g(p) )\ \exp \left( \int dq\, g(q) c_b(q) \frac{\delta }{\delta c_b(q)}  \right) \frac{\delta }{\delta c_a(p)}    
\end{eqnarray}
where $c_b(q)$ is a Grassmann variable and $g(q)$ is a real function of $q$. 
From Eq.~\ref{EqPiSymmetry} and \ref{EqChiSymmetry} we have
\begin{eqnarray}
   \tilde{c}^{*\,a}(\vec{p}) \, \eta =  \omega_p \, dx\, \eta\, \tilde{c}^*_a (\vec{p}),\label{EqnB2} 
 \\
    (\tilde{c}_a(\vec{p}))^\dag \, \eta =  \frac{1}{\omega_p\,dx}  \eta\, (\tilde{c}_a(\vec{p}))^\dag. \label{EqnB3}
\end{eqnarray} 
\black{Taking the complex conjugate of Eq.~\ref{EqnB3}, noting that
  $(\tilde{c}^*(\vec{p}))^\dag = \delta/\delta \tilde{c}(\vec{p})$, setting $\exp(g(p))=(dx\,\omega_p)^{-1}$, and  recognizing that $\eta = \eta^*$, we verify that}
\begin{equation}
        \eta  =  \exp \left( - \int d^3\vec{q}\, \log(dx\,\omega_q) c_b(\vec q) \frac{\delta }{\delta c_b(\vec q)}  \right)
\end{equation}
which is equivalent to Eq.~\ref{EqEtaDef}.
The normalization of $\eta$ is not uniquely determined independent of the normalization of the states.

Alternatively following Mostafazadeh \cite{Mostafazadeh_2002}, we can also find an 
under-constrained definition of $\eta$ through the eigenvectors of $H$ and the eigenvectors of $H^\dag$. 
Placing the eigenvectors of $H$ as columns in the matrix $V$, and the eigenvectors of $H^\dag$ as columns in the matrix $Y$, then
up to degeneracies and zero eigenvalue ambiguities, 
\begin{equation}
    \eta= Y V^{-1}.
    \label{EqEtaAlt}
\end{equation}  
Eq.~\ref{EqEtaAlt} does not depend on finding a mode decomposition.
In contrast, Eq.~\ref{EqEtaDef} is not under-constrained and is compatible with the constrained elements of Eq.~\ref{EqEtaAlt}.

\section{Derivatives, Translations, and Momentum on a Torus}
\label{SecTorusDerivatives}

Momentum on a fermionic lattice model has several subtleties.  It depends on the map between the coordinates and the qubits.  Derivatives which cross a cyclic boundary pick up a state dependent phase.  To address these, first we define the map between qubits and coordinates, then we define the translation operators.  Using the map and the translation operators, we can then define the derivatives and momentum operators in a general manner.
Then we will form an operator which exchanges spatial directions.  
This will enable us to define the derivatives in any number of dimensions.

Consider a lattice with  $N_S$ spinor indices, and $(N_X,N_Y,N_Z)$ points in the $(X,Y,Z)$ spatial directions.
Our first step is to assign the $r^{\texttt{th}}$ qubit to the spatial coordinates $dx\,(i,j,k)$  and the spinor index \black{$a\in\{0, \ldots  , N_S-1\}$} through:
\begin{equation}
   r=i\, N_Y N_Z N_S + j\,  N_Z N_S + k\, N_S + a.
   \label{EqrBitDef}
\end{equation}
\begin{widetext}
The spinor and spatial qubits  can also be represented as \black{the multi-qubit state vector}:
\begin{equation}
  \Bigl|\,  \underbrace{   \underbrace{
  \underbrace{\texttt{00}}_{k = N_{Z}-1} \;  \ldots 
  \underbrace{\texttt{00}}_{k = 0} \;}_{j=N_Y-1} 
  \ldots  \underbrace{
  \underbrace{\texttt{00}}_{k = N_Z-1} \; \ldots
  \underbrace{\texttt{00}}_{k = 0} \;}_{j=0} 
  }_{i=N_X-1} \; \ldots \ldots \; \underbrace{   \underbrace{
  \underbrace{\texttt{00}}_{k = N_{Z}-1} \;   \ldots 
  \underbrace{\texttt{00}}_{k = 0} \;}_{j=N_Y-1} 
  \ldots   \underbrace{
  \underbrace{\texttt{00}}_{k = N_Z-1} \; \ldots
  \underbrace{\texttt{00}}_{k = 0} \;}_{j=0} 
  }_{i=0}\,\Bigr\rangle.
  \label{EqStateDefInMultiDims}
\end{equation}
The coordinates $(i,j,k)$ are zero indexed, and we display the case with $N_S=2$.
\end{widetext}

\begin{widetext}

\begin{figure}
\centering
\begin{tabular}{ccc}
\hspace{-0.5in}
\includegraphics[width=3.75in,height=0.9in]{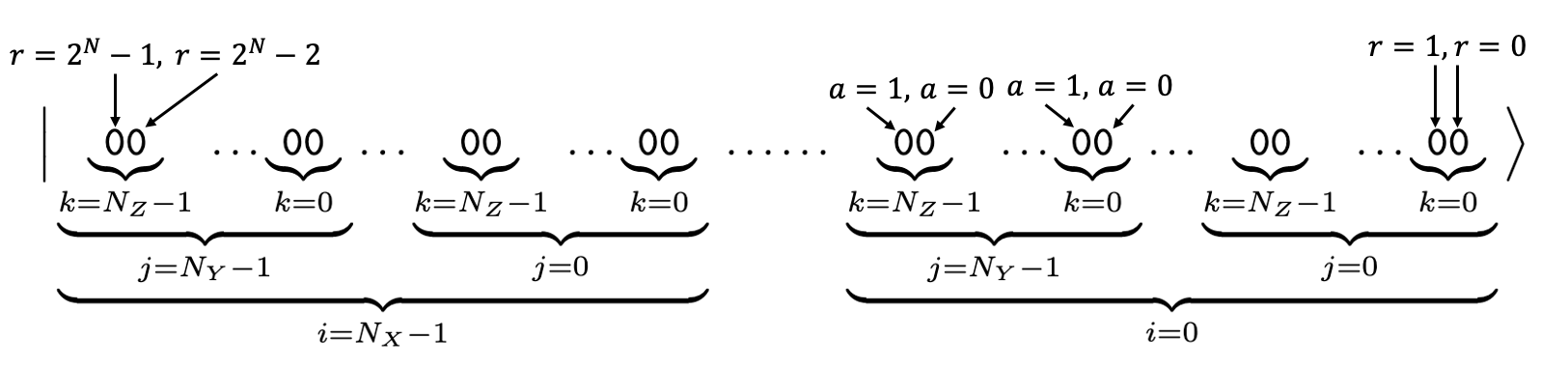} &
\hspace{0.1in} {}&
\includegraphics[width=3.75in,height=0.9in]{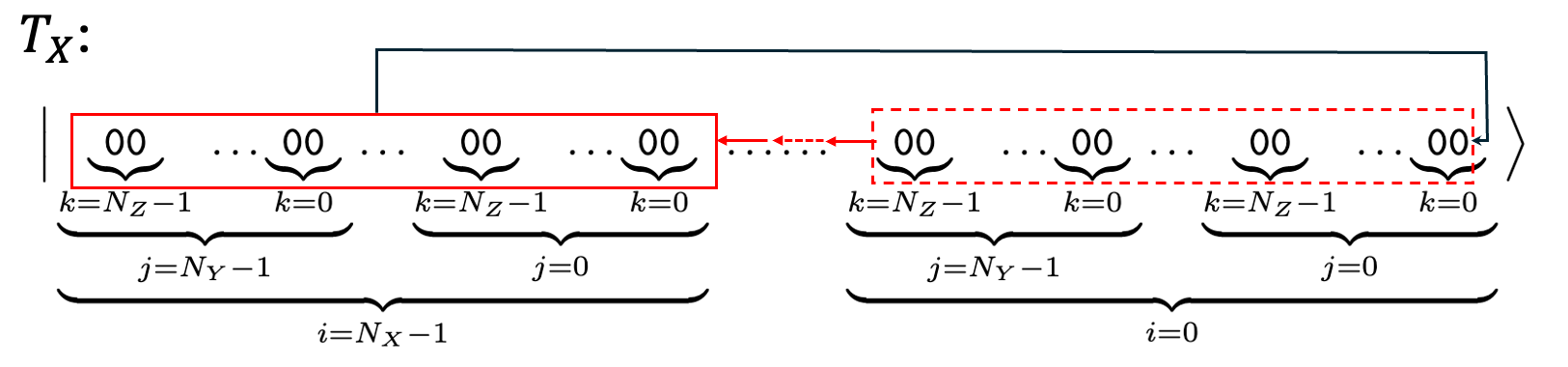} \\ 
\hspace{-0.5in}
\includegraphics[width=3.75in,height=0.9in]{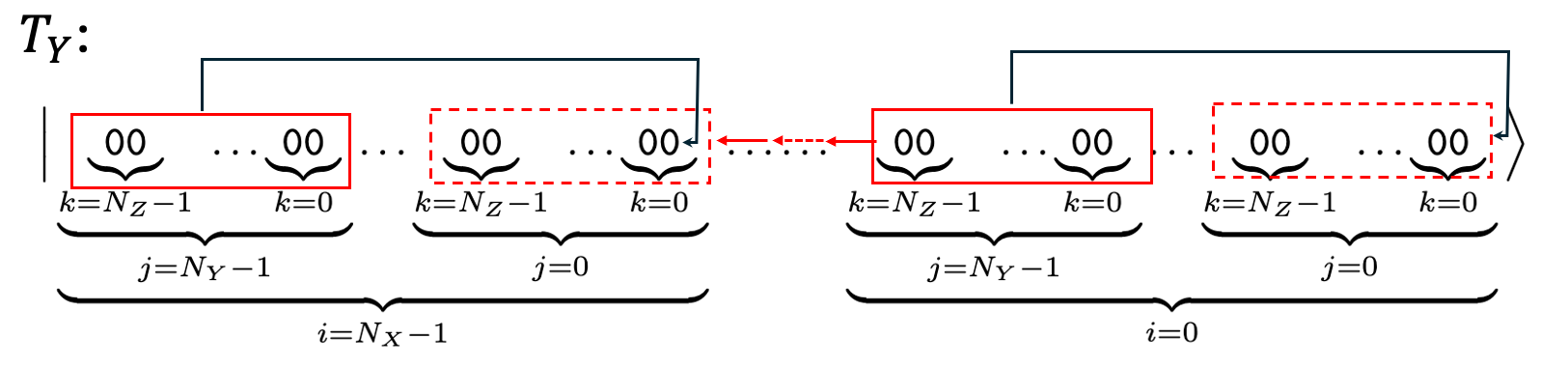} &
\hspace{0.1in} {}&
\includegraphics[width=3.75in,height=0.9in]{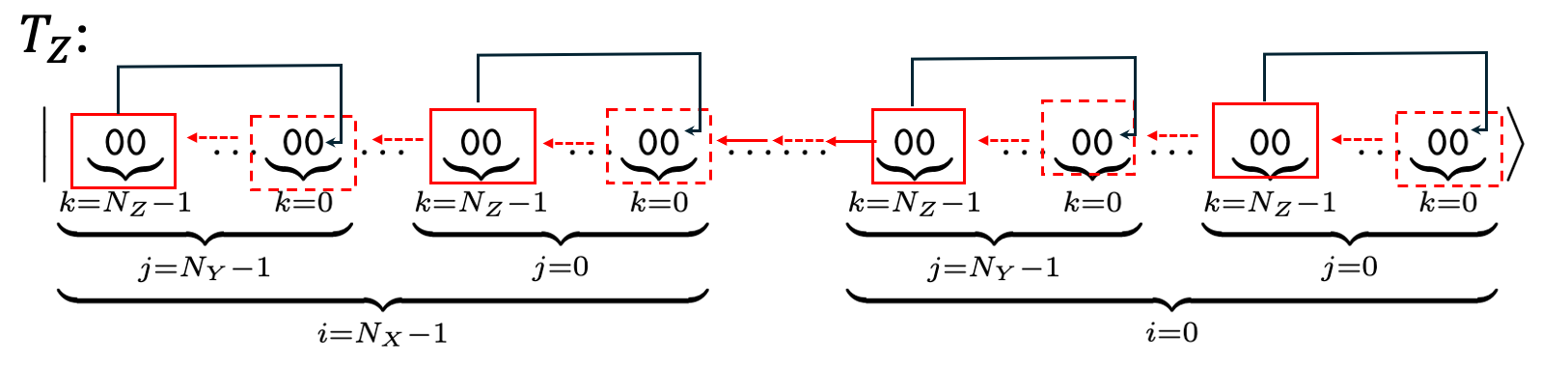}
\end{tabular}
\caption{
(top left) Spinor and spatial qubits indexing. 
Schematic of block movements of qubits for the one site translations:\newline
(top right) $T_X$ (\ref{EqTX}),
(bottom left) $T_Y$(\ref{EqTY}), 
(bottom right) $T_Z$ (\ref{EqTZ}).
}\label{fig:C2:TX:TY:TZ}
\end{figure}

\end{widetext}

There are $N=N_X\,N_Y\,N_Z\,N_S$ qubits and therefore $2^N$ states.
We start with the operator that results in a 
cyclic left-shift translation by $1$-qubit of the $N$ qubits given by
\begin{equation}
    (U_{(L;N)})_r^{\ c} = 
    \delta^c_{ 2\,r\Mod{2^N} } + {\text{floor}}\left(\frac{2\,r}{2^N}\right)
    \label{EqShiftULDef}
\end{equation}
where $\text{floor}(a/b)$ removes the fractional part of  $a/b$.
The indices $r$ and $c$ run from $0$ to $2^N-1$.
The left-shift cyclic operator defined in Eq.~\ref{EqShiftULDef} acts on a state of $N$ bits as follows:
\begin{equation}
    U_{(L;N)}  |\, x_{N-1} \, x_{N-2} \,  \ldots  x_{1}\, x_{0} \rangle = 
     |\,  x_{N-2} \,  \ldots  x_{1}\, x_{0} x_{N-1} \rangle.
\end{equation}
The unitary translation operators that shift the lattice by one site in the $(X,Y,Z)$ directions are given by
\begin{eqnarray}
T_X=(U_{(L;N)})^{N_Y\,N_Z\,N_S},
\label{EqTX}
\\
    T_Y = \bigotimes^{N_X}  ( U_{(L;N_Y N_Z N_S)} )^{N_Z\,N_S},
    \label{EqTY}
\\
    T_Z = \bigotimes^{N_X N_Y}  (U_{(L;N_Z N_S)})^{N_S}.
    \label{EqTZ}
\end{eqnarray}
The $\bigotimes^{N_X} $ represent tensor products of the the argument for $N_X$ times.
In Eq.~\ref{EqTX}, we shift to the left by $N/N_X=N_Y N_Z N_Z$.
This is done by raising the $U_{L;N}$ to the power $N/N_X$. This result acts so that every element of $i=0$ is moved to $i=1$, and every element of $i=1$ is moved to $i=2$.  Because it is cyclic, every element of $i=N_X-1$ will be shifted to $i=0$.
In Eq.~\ref{EqTY}, we shift cyclically within each block of a fixed $i$.  For this reason, the shift operator used is $U_{L;N/N_X}$.
Again we shift so that every element of $j=0$ is moved to $j=1$, and so forth.  
This means we have to shift by $N_X N_S$ bits which is why the shift operator is raised to this power.
The $\bigotimes^{N_X}$ ensures that this shift is done for each grouping of $i=0, i=1, \ldots i=N_X-1$. 
The pattern occurs again in Eq.~\ref{EqTZ}.
This time we are shifting by $N_S$ (which we have chosen as $2$ 
the depiction above).
Our shift operator acts on regions of constant $i$ and constant $j$ and sends every $k=0$ to $k=1$, $k=1$ is moved to $k=2$, etc.
\black{The  movements of $X$, $Y$ and $Z$  blocks of qubits for the corresponding one site translations $T_X, T_Y, T_Z$ are illustrated in Fig.~\ref{fig:C2:TX:TY:TZ}. }
This completes the definition of translation operators on the lattice.

Next, we define the derivatives on the lattice  to  preserve cyclic translational invariance.
Derivatives for fermions on a ring or a torus require corrections when a derivative spans the boundaries where the fermions wrap. 
When implemented correctly, the Laplacian operator $\nabla^2$ commutes with with the shift operator in each direction.

Taking derivatives across the boundary naively does not necessarily commute with the translation operator.
When a derivative crosses a boundary, the result should be the same as if it were `shifted' over, the derivative taken on a region without a boundary, and then the result is shifted back.
We can accomplish this by using the translation operators \black{on the 
innermost block of qubits representing $Z$, and then using} $T_Z$ to shift the bits to one side of the boundary, take the derivative, and then shifting the bits back to their original positions.  
\black{We will accomplish the derivatives in $X$ and $Y$ directions by using an operator to exchange their associate outer blocks of qubits with the innermost block of qubits representing $Z$.}

We demonstrate this with the second-order derivative in \black{the $Z$} dimension including an internal spinor index, where 
\black{we suppress the indices for the other two dimensions.} 
The results generalize to $2$ and $3$ spatial dimensions using the dimensional exchange operators in Eq.~\ref{EqOpDimensionExchange}. 
We use the wrap function $w(k)=k \Mod {N_Z}$ to wrap around the $Z$-dimension of the ring or torus.
The $1$D Laplacian term is discretized as $ \int d^3x \, \chi \gamma^0 \partial_z^2 \chi \rightarrow$:
\begin{eqnarray}
&=&  \sum_{k=0}^{N_Z-1}  \frac{\chi_a(k) (\tilde{\gamma}^0)^{ab}(\chi_b(w(k+1)) + \chi_b(w(k-1)) -2\chi_b(k))}{dz^2} \nonumber \\
&=& 2 \sum_{k=0}^{N_Z-1} \,    \left(  \frac{ \chi(k) (\tilde{\gamma}^0) \chi(w(k+1))  - \chi_a(k) (\tilde{\gamma}^0) \chi(k)}{dz^2} \right), \label{EqnC8}
\end{eqnarray}
where in the second line we have reordered the terms so that we only have to consider wrapping around one side of the boundary.
We also set $dz=dx=dy$ as the lattice spacing.
\black{Note that in Eq.~\ref{EqnC8} the
 sum over the $Z$-dimension index $k$ is also summed over each point $\{X,Y\}\leftrightarrow\{i,j\}$  in the computational grid (where we have suppressed these latter indices and sums in Eq.~\ref{EqnC8}).}
In this simplified case, the translation operator for one point shifts over by the number of internal spinor indices as defined in Eq.~\ref{EqTZ}.
Now there is one entry where $k=w(N_Z)=0$ where the wrap function is invoked.
We can address that one term individually, so we break out the last entry as $\int d^3 x \, \chi \tilde{\gamma}^0 \partial_z^2 \chi \rightarrow$:
\begin{eqnarray}
& =& 2 \sum_{k=0}^{N_Z-2} \,    \left(  \frac{ \chi_a(k) (\tilde \gamma^0)^{ab} \chi_b(w(k+1))  - \chi_a(k) (\tilde \gamma^0)^{ab} \chi_b(k)}{dz^2} \right) \nonumber \\ 
& & 
-  \left(  \frac{   \chi_a(N_Z-1) (\tilde \gamma^0)^{ab} \psi_b(N_Z-1)}{dz^2} \right) \nonumber \\
& & 
+ \left(  \frac{ \chi_a(N_Z-1) (\tilde \gamma^0)^{ab} \chi_b(0)  }{dz^2} \right), \label{EqnC10}
\end{eqnarray}
\black{where we have again suppressed the  $\{X,Y\}\leftrightarrow \{i,j\}$ indices and sums in Eq.~\ref{EqnC10}}.
However, the states on which $\chi(0)$ act may have the wrong sign relative to $\chi(N_Z-1)$ after wrapping around.
To  correct for the wrapping, 
we use the translation operator to shift by one point to perform the wrapping correctly:
\begin{eqnarray}
      \chi_a(N_Z-1) (\tilde \gamma^0)^{ab} \chi_b(0)  \rightarrow \hspace{2cm} \nonumber \\ 
      (T_Z)^\dag \chi_a(N_Z-2) (\tilde \gamma^0)^{ab} \chi_b(N_Z-1) T_Z.
\end{eqnarray}
In this case the derivative is performed between $N_Z-2$ and $N_Z-1$ on one side of the boundary, and the the shift operator moves the operator back over the boundary.
We substitute this result in for the last  entry  such that the Z component of the Laplacian becomes $\int d^3x \, \chi \gamma^0 \partial_z^2 \chi \rightarrow$
\begin{eqnarray}
& = &  2 \sum_{k=0}^{N_Z-2} \,    \left(  \frac{ \chi_a(k) (\gamma^0)^{ab} \chi_b(w(k+1))  - \chi_a(k) (\gamma^0)^{ab} \psi_b(k)}{dz^2} \right) \nonumber \\ 
& & 
-  \left(  \frac{   \chi_a(N_Z-1) (\gamma^0)^{ab} \psi_b(N_Z-1)}{dz^2} \right) \nonumber \\
& & 
+ \left(  \frac{ (T_Z)^\dag \chi_a(N_Z-2) (\gamma^0)^{ab} \chi_b(N_Z-1) T_Z }{dz^2} \right), 
\label{EqnC12}
\end{eqnarray}
 \black{again suppressing the  $\{X,Y\}\leftrightarrow \{i,j\}$ indices and sums in Eq.~\ref{EqnC12}}.
This defines the derivative operator that commutes with the $T_Z$ translation operator.

\black{In order to perform the $X$ and $Y$ component of the Laplacian, we will first map them into the $Z$ component of the Laplacian described above.}
This is most easily done when $N_X=N_Y=N_Z$.
\black{Consider the $Y$ component of the Laplacian.}
We begin with a matrix that exchanges the \black{ qubits } that describe $Z$ and $Y$ based on Eq.~\ref{EqrBitDef}. 
The matrix
\begin{eqnarray}
    B^r_{\ r'} = \hspace{2.9in} \label{EqBrr} \\ 
    \delta^r_{a+k N_S  + j N_S N_Z + i N_X N_Z N_Y}  \delta_{r'}^{a + j N_S  + k N_S N_Z + i N_X N_Z N_Y} \nonumber 
\end{eqnarray}
where we sum over $a,i,j,k$. 
Notice that Eq.~\ref{EqBrr} simply exchanges $k \leftrightarrow j$.  
As such $B^2 = \unit$.
The matrix $B$ now acts on the vector defined in Eq.~\ref{EqStateDef} to exchange the location of the bits such that $Z$ and $Y$ are interchanged \black{ on the state labels}:
\begin{equation}
    |\vec{x}'\rangle = | B \vec{x} \rangle.
\end{equation}
Here $|\vec{x'}\rangle$ is the state onto which $|\vec{x}\rangle$ maps when one exchanges $Z$ with $Y$.
Using this we can define the operator which exchanges $Y$ and $Z$ as
\begin{equation}
    G=\sum_{\vec{x}} |  B \vec{x} \rangle \langle \vec{x} |.
    \label{EqOpDimensionExchange}
\end{equation}
The operator $G$ is unitary and satisfied $G^2=\unit$.
From here we can define the $Y$ second derivative in terms of the $Z$ second derivative by:
\begin{equation}
    \int d^3 x \, \chi \gamma^0 \partial_y^2 \chi =  G \left( \int d^3 x \, \chi \gamma^0 \partial_z^2 \chi \right) G.
\end{equation}
The second derivative on $X$ can by built with an analogous approach. The overall Laplacian is then the sum of the operators with second derivatives in each direction.

With the translation and derivative operator defined, we can now define the momentum   operator as the log of the translation operator:
\begin{equation}
\hspace{-0.2in}
    (P_X,P_Y,P_Z) = ( \frac{-i\,\log T_X}{dx}, \frac{-i\,\log T_Y}{dx}, \frac{-i\,\log T_Z}{dx}).
\end{equation}
Because the lattice may be occupied by neighboring sites, this is not simply the exponentiation of a $\pi_\chi \partial_\mu \chi$ term. 

The definition of momentum in terms of the translation operator makes solving for the eigenvalues straightforward. 
Because these translation operators satisfy $(T_X)^{N_X}=1$, so the eigenvalues will be the $N_X$ roots of unity given by powers of $\exp(i\,2\pi /N_X)$. 
Likewise $(T_Y)^{N_Y}=1$ and $(T_Z)^{N_Z}=1$.
Therefore we can see that the momentum eigenvalues $\vec{P}$ are given by increments of $(n_x\,\frac{2\pi}{dx\,N_X},n_y\,\frac{2\pi}{dx\,N_Y}, n_z\,\frac{2\pi}{dx\,N_Z})$ where $(n_x,n_y,n_z)$ are triplets of integers indexing the modes.

In $1$D, the Laplacian \black{$\nabla^2_T$} that commutes with the translation operator can equivalently be expressed as
\begin{eqnarray}
   (\nabla_T^2)^{j\,k} = \frac{\delta^{j,{w(k+1)}}\Upsilon_{(j,N_X-1)} + \delta^{j,{w(k-1)}}\Upsilon_{(j,0)} - 2\, \delta^{j,k} }{dx^2}, \qquad\quad
\end{eqnarray}
where $w(j)=j \Mod {N_X}$ and
\begin{equation}
   \Upsilon_{(a,b)} = \left\{ \begin{array}{ccc}
      \text{If}\ a=b & \text{then}\   \Gamma_{N+1} \\
      \text{If}\  a\neq b & \text{then}\ \unit
   \end{array}    \right. 
\end{equation}
The $\Gamma_{N+1}$ is a generalization of a chiral gamma matrix given by
\begin{eqnarray}
    \Gamma_{N+1} & = & -(-i)^N \prod_{j=0}^{N-1} \Gamma_j \\
    & = & (-1)^{\sum_{j=0}^{N_X-1} i\,\pi^B(j) \xi_B(j)}.
    \label{EqProjectors}
\end{eqnarray}
This effectively divides the spectrum up into \black{two sectors: one where $\Gamma_{N+1}$ gives $1$ which look like periodic boundary conditions, and one where $\Gamma_{N+1}$ gives $-1$ which look like antiperiodic boundary conditions.}
The underlying system is cyclically translation invariant, but the numerical manifestation has these periodic and antiperiodic boundary conditions.  

To aid in solving the $1+1$D case exactly, we project onto these two spaces $P_{(P)}=\frac{1}{2}(\unit + \Gamma_{N+1})$ and  $P_{(A)}=\frac{1}{2}(\unit - \Gamma_{N+1})$.  From here we can express the translationally invariant Laplacian in terms of the periodic \black{$\nabla^2_P$}, and antiperiodic \black{$\nabla^2_A$}, boundary condition Laplacians via the identity
\begin{equation}
    \nabla^2_T =  \nabla^2_P P_{(P)} +  \nabla^2_{A} P_{(A)}
\end{equation}
where 
\begin{equation}\label{delsquaredP}
    (\nabla^2_P)^{j,k} =  \frac{ \delta^{j,w(k-1)} +  \delta^{j,w(k+1)} - 2 \delta^{j,k}  }{dx^2} ,
\end{equation}
and
\begin{equation}\label{delsquaredA}
     (\nabla^2_A)^{j,k} =  \frac{ \delta^{j,w(k-1)} (-1)^{\delta_{c,0}} +  \delta^{j,w(k+1)} (-1)^{\delta_{c,N_X-1}} - 2 \delta^{j,k}  }{dx^2}. 
\end{equation}
The operator $\nabla^2_P$ has eigenvalues
\begin{equation}
    -k_P^2(n) = \frac{-4}{dx^2} \sin^2 \left( \frac{n\,\pi}{N_X} \right) ,
\end{equation}
and $\nabla^2_A$ has eigenvalues
\begin{equation}
  -k^2_A(n) = \frac{-4}{dx^2}\sin^2\left( \frac{(n+\frac{1}{2})\pi}{N_X} \right). 
  \label{EqkSqADef}
\end{equation}
These $(A)$ and $(P)$ modes are used in the mode decomposition in Sec.~\ref{Sec1D6PModel}.

\bibliography{NoDoublers}

\end{document}